\documentclass[a4paper,fleqn]{cas-sc}
\usepackage[numbers,square]{natbib}
\usepackage[T1]{fontenc}
\usepackage[utf8]{inputenc}
\usepackage{amsmath,amsfonts}
\usepackage{microtype}
\usepackage{graphicx}
\usepackage{subcaption}
\usepackage{booktabs}
\usepackage{placeins}
\usepackage{titletoc} %
\usepackage{cleveref}
\usepackage{hyperref}
\usepackage{url}

\makeatletter
\def\fps@figure{htbp}
\def\fps@table{htbp}
\makeatother

\begin{document}
\shorttitle{Fault Detection in Solar Thermal Systems}

\shortauthors{Ebmeier et al.}

\title [mode=title]{Fault Detection in Solar Thermal Systems using Probabilistic Reconstructions}

\author[1]{Florian Ebmeier}[%
  orcid=0000-0002-1518-9593,
]
\cormark[1]
\ead{florian.ebmeier@uni-tuebingen.de}
\credit{Writing – original draft, Software, Data curation, Methodology, Investigation}

\author[1]{Nicole Ludwig}[
  orcid=0000-0003-3230-8918
]
\credit{Supervision, Funding acquisition, Writing – review \& editing, Methodology, Conceptualization}

\author[1]{Jannik Thuemmel}[
  orcid=0009-0004-4209-6239
]
\credit{Software, Writing – review \& editing, Methodology, Investigation}

\author[1,2]{Georg Martius}[
  orcid=0000-0002-8963-7627
]
\credit{Funding acquisition, Writing – review \& editing, Methodology, Conceptualization}

\author[1]{Volker H. Franz}[
  orcid=0000-0002-8457-0611
]
\credit{Supervision, Funding acquisition, Writing – review \& editing, Methodology, Conceptualization}

\cortext[1]{Corresponding author}

\affiliation[1]{organization={University of Tübingen},
  addressline={Geschwister-Scholl Platz},
  city={Tübingen},
  postcode={72072},
  state={Baden-Württemberg},
  country={Germany}}

\affiliation[2]{organization={Max Planck Institute for Intelligent Systems},
  addressline={Max-Planck-Ring 4},
  city={Tübingen},
  postcode={72076},
  state={Baden-Württemberg},
  country={Germany}}
\begin{abstract}
  Solar thermal systems (STS) present a promising avenue for low-carbon heat generation, with a well-running system providing heat at minimal cost and carbon emissions. However, STS can exhibit faults due to improper installation, maintenance, or operation, often resulting in a substantial reduction in efficiency or even damage to the system. As monitoring at the individual level is economically prohibitive for small-scale systems, automated monitoring and fault detection should be used to address such issues. Recent advances in data-driven anomaly detection, particularly in time series analysis, offer a cost-effective solution by leveraging existing sensors to identify abnormal system states.

  Here, we apply probabilistic reconstruction-based anomaly detection to the publicly available PaSTS dataset of operational domestic STS, which features real-world complexities and diverse fault types. We combine an LSTM-based Variational Autoencoder with heteroscedastic uncertainty estimation and evaluate it exclusively on systems unseen during training.
  Our experiments show that reconstruction-based methods detect faults in domestic STS both qualitatively and quantitatively. Our model achieves the best generalisation to unseen systems among all evaluated methods, while a rescaled PCA reconstruction baseline remains competitive on pooled threshold-free metrics. An ablation shows that heteroscedastic uncertainty estimation drives the gain in cross-system generalisation.
  Finally, we discuss the engineering overhead required to unlock these improvements and make a case for simple deep learning models.
\end{abstract}
\begin{keywords}
  Solar thermal systems \sep Fault detection \sep Anomaly detection \sep Time series \sep Deep learning \sep Uncertainty estimation
\end{keywords}
\maketitle
\section{Motivation}
\label{sec:motivation}
Domestic solar thermal systems (STS) present a promising approach to decarbonising the energy system. They require minimal natural resources, do not add strain to the electrical grid, and can be more space-efficient than photovoltaic systems. However, adoption has been hampered by low yields due to poor monitoring and undetected faults \cite{paerisch2006quality, paarisch2007wissenschaftlicher}. Monitoring these systems can be prohibitively expensive if it involves additional sensors, complex system modelling, and continuous oversight by human experts. With increased digitalisation, substantial sensor data are now available, creating an opportunity to develop data-driven fault and anomaly detection methods that reduce costs and human supervision.

Data-driven time series anomaly detection (TSAD) has attracted considerable attention in machine learning. Among the prevailing paradigms, reconstruction- and forecasting-based methods are particularly popular, and recent years have seen a surge in the application of cutting-edge deep learning (DL) architectures to this task \cite{an_variational_2015, audibert_usad_2020, deng_graph_2021, su_robust_2019, park_multimodal_2018, tuli2022tranad}.
To our knowledge, no system applies state-of-the-art deep learning anomaly detection frameworks to real-world domestic STS. This is particularly severe when considering the variety of system realisations in operational domestic STS. Generalisation remains the key challenge: to build robust models that can adapt to the variety of real-world system designs, operational states, and environmental factors common in domestic solar thermal setups.
Furthermore, concerns have been raised regarding the soundness of experimental protocols and benchmarking practices in many DL studies \cite{wu_current_2021, kim_towards_2022,sarfraz_position_2024}. For example, \citet{sarfraz_position_2024} demonstrate that simple linear methods such as principal component analysis (PCA) can outperform recent, more complex DL models; thus challenging the presumed superiority of advanced architectures.

The finding that PCA substantially outperforms DL models in TSAD is surprising, as there is no theoretical basis for this shortcoming of DL models.
\citet{RolinekZietlowMartius:VAERecPCA} show that under standard training Variational Autoencoders (VAEs) locally align their latent representation to the principal directions of the data, occupying the same representational regime as PCA. There is an even stronger connection if we consider that a linear Autoencoder trained with a reconstruction loss recovers PCA directly \cite{baldi1989neural}. We would therefore expect VAE performance to be on par with PCA, with the differences between the models attributed to the non-linear decoder and expected to be small.
Thus, we hypothesise that this performance gap is not due to the inherent shortcomings of DL models, but rather to the added complexity of optimising overly complicated DL models. Hence, we focus on simpler methods that require little optimisation while keeping the strengths of DL models. More specifically, we show that probabilistic reconstruction-based methods make fault detection in STS feasible, and that explicit heteroscedastic uncertainty estimation is a simple yet effective method to enhance fault detection accuracy.

Our experiments on the PaSTS dataset \cite{ebmeier_pasts_2024} will explore these hypotheses, offering insights into how data-driven techniques can be effectively integrated into domestic STS monitoring workflows.

The structure of the paper is as follows: First, we briefly introduce and discuss the relevant literature on fault detection in STS and TSAD, identifying the research gap and our research contribution (Section 2). We then introduce the methods we use to close this gap (Section 3). Following this, we introduce the Experiments, in particular the dataset and experimental setup, and then show our results (Section 4). In Section 5, we conclude and discuss the key takeaways of the study as well as possible improvements for the future.

This paper demonstrates that reconstruction-based anomaly detection methods are both qualitatively and quantitatively successful in fault detection in STS. Our method achieves the best generalisation to unseen systems among all evaluated methods, including simple statistical baselines, such as PCA, as well as both simple and complex DL baselines.
Additionally, uncertainty estimation further improves detection, with heteroscedastic uncertainty estimation leading to the best performance without requiring additional hyperparameters. To our knowledge, the contribution of heteroscedastic uncertainty estimation to detection performance has not previously been isolated in the TSAD literature. Finally, we assess the engineering and optimisation cost at which these gains are obtained, relative to simple baselines.

\section{Background}
\subsection{Fault detection in solar thermal systems}
Fault detection in STS has been approached in various ways, as discussed in several works \cite{de_keizer_review_2011, faure_fault_2020}. These approaches can be broadly categorised into rule-based and expert systems, modelling or parity-space approaches, and data-driven approaches, with some overlap between these categories. In practice, operational fault detection often combines elements of these approaches, leveraging domain expertise and automated detection tools to identify and diagnose failures.

\textbf{Rule-based and expert systems}. Rule-based and expert systems detect faults by checking if a system's behaviour deviates from predefined rules. These rules are crafted using detailed system knowledge to model specific failure cases individually tailored to the system type. As a key advantage, they often directly classify the specific fault whenever a rule is violated. A simple example is sensor fault detection: if a sensor produces highly implausible outputs, such as jumping several orders of magnitude outside its dynamic range, it is considered a faulty sensor. However, this approach only detects anticipated failures encoded in the rules. Any anomalies or failures not considered during rule creation will go undetected. Expert and rule-based systems have been employed, for example, in the IP-Solar project \cite{droscher2009modular, ohnewein2010ip, holter2012development}, the FeDet research project \cite{kuethe2011implementation, schmelzer2015fedet, georgii2017flexible}, as well as the Solar-Check project \cite{schmelzer2021solarcheck,georgii2022digital}.

\textbf{Modelling approaches}. Modelling approaches, including parity-space and physics-based methods, focus on replicating system dynamics through simulations or analytical equations. This is done either through simulation tools, simulating the expected behaviour of the system \cite{de_keizer_simulation-based_2013}, or by applying physical equations to estimate the yield of the collector loop directly \cite{paerisch2006quality, paarisch2007wissenschaftlicher}. Large deviations from the estimations are then flagged as faults. These approaches thus detect any unexpected faults or service interruptions without specifying the fault type beforehand and are more robust to unexpected behaviour. However, unlike the rule-based system, the faults are not immediately classified but require a secondary approach or manual system analysis. Additionally, modelling approaches often require detailed knowledge of the specific system design and supplementary sensors, further constraining their applicability.

\textbf{Data-driven approaches}. Data-driven approaches do not rely on explicit physics-based models or rules; instead, they learn patterns directly from data. Early examples in STS used small, simple neural networks to estimate sensor readings, relying on residuals between predictions and measurements to identify anomalies \cite{kalogirou_development_2008,kalogirou_artificial_2014, ferreiro_garcia_monitoring_2014}.
Some recent works have extended these concepts to recurrent neural network architectures and slightly more complex, modern deep learning architectures \cite{correa_jullian_assessment_2019}. Additionally, there has been work that uses operational data from large STS to train tree-based models, similar to previous work, which uses tree regression models to estimate sensor values based on both historical data and data from highly correlated sensors \cite{feierl_fault_2023}.

\textbf{Data sources and availability}
All of these works have been trained on nominal data and then evaluated on both nominal and faulty data. The data sources of these works have been simulations and pilot projects, which have very controlled setups and defined layouts. Typically, the faults detected were either induced by the pilot project operators or programmed into the simulation. Additionally, the models employed were for a single specific STS layout. Therefore, they may struggle to generalise across different installations and sensor layouts, which is frequently encountered in domestic STS, where configurations vary widely across systems. In contrast, we recently published a publicly available dataset of operational domestic STS, featuring a wide variety of real-world system realisations and faults \cite{ebmeier_pasts_2024}.

\subsection{Time Series Anomaly Detection}
\label{sec:tsad}
\textbf{Deep learning in TSAD}
Time series anomaly detection in deep learning initially centred on Autoencoders or forecasting models, building on their success in related tasks, such as computer vision and speech recognition. Early efforts focused on embedding modern architectures into these frameworks, including VAEs with long short-term memory (LSTM)-based encoding and decoding \cite{park_multimodal_2018}, time-dependent latent spaces, and normalising flows \cite{su_robust_2019}. Seeking further improvements, subsequent studies explored more complex strategies, such as adversarial training \cite{audibert_usad_2020}, novel transformer-based designs \cite{tuli2022tranad}, modified attention mechanisms \cite{xu_anomaly_2022}, graph neural networks \cite{deng_graph_2021}, and contrastive learning \cite{darban_carla_2025}.
These approaches all assign anomaly scores to individual points or sections of a time series, which can then be either directly analysed or thresholded for automatic classification. While there are many different ways to generate these anomaly scores based on the used architecture, forecasting- and reconstruction-based methods usually take a distance in data space, such as the mean-squared error, as an anomaly score.

\textbf{Critique of TSAD literature}
However, recent investigations have highlighted fundamental issues in the TSAD literature. Commonly used benchmarks often contain noisy or mislabelled faults, some of which are trivial or impossible to detect, thus skewing model performance. Furthermore, other biases are present in commonly used datasets, such as the run-to-failure bias and unreasonable anomaly densities \cite{wu_current_2021,NEURIPS2024_c3f3c690,sarfraz_position_2024}. There are also no clearly established guidelines for analysis and preprocessing. For example, \citet{sarfraz_position_2024} reported some studies using 127 and other studies using 112 features on the WADI dataset, which itself reports to only have 125 features.
In addition, the metrics favoured in the field, specifically the point-adjusted F1 score, tend to reward continuous or noisy predictions, with some studies showing almost perfect results for randomly chosen anomaly scores \cite{kim_towards_2022}. Other approaches to replace problematic metrics, e.g., segment-based alternatives \cite{tatbul_precision_2018}, have either failed to achieve large-scale adoption or introduced new problems. The same holds for proposed changes to benchmark workflows.

\textbf{Simple models outperform complex models}
When restricting the analysis to the least problematic benchmarks and correcting for evaluation flaws, recent publications have found simpler methods to consistently outperform more elaborate models. \citet{sarfraz_position_2024} found that for all their tested DL architectures, a simplified forecasting method outperformed the established DL method. Furthermore, both these simple methods as well as the complex DL architectures were outperformed by a PCA-based reconstruction method using simple error rescaling.
Other research, such as the recent benchmark dataset by \citet{NEURIPS2024_c3f3c690}, also found classical machine learning methods as well as simpler DL methods outperform complex DL methods both for univariate and multivariate time series. This performance gap was especially pronounced in the univariate case.

\subsection{Research Gap}

Despite the abundance of raw sensor data, previous research on fault detection in STS has been constrained to either synthetic or proprietary datasets, as manufacturers are often unwilling or unable to share operational data. Only the recently published PaSTS dataset \cite{ebmeier_pasts_2024} is publicly accessible for operational domestic STS. This dataset allows us to extend the fault detection literature on STS by a data-driven approach that leverages the insights of the recent DL--TSAD literature.

As we aim to stick closely to the established workflow from \citet{ebmeier_pasts_2024}, we introduce several constraints. First, this is operational data from heterogeneous, non-standardised domestic STS, requiring methods with strong generalisation. To ensure this, we validate and test only on entirely unseen systems.
Second, we focus on post-hoc system analysis, rather than real-time control. This enables purely offline evaluation and aligns with existing fault-diagnosis workflows. As these workflows interpret anomalies within a broader system context, we analyse full-day sequences rather than shorter segments, localising anomalies at this coarser granularity.

We adopt a semi-supervised anomaly detection setup, common in TSAD, where training occurs only on nominal data. In this semi-supervised setup, the model is trained exclusively on data representing nominal system behaviour. Anomalies are then identified as deviations from this learned distribution during inference. This approach aligns with practical constraints, where labelled faults are scarce, unavailable or unreliable.
We implement a straightforward LSTM-based VAE, informed by the recent TSAD literature. Building on findings that error rescaling improves detection by implicitly modelling uncertainty, we extend the VAE with probabilistic reconstructions, comparing homoscedastic and heteroscedastic uncertainty estimation.

\section{Methods}
\subsection{Likelihood-based anomaly detection}
\label{sec:reconstruction_ad}
\textbf{Reconstruction-based anomaly detection}
Reconstruction-based anomaly detection relies on deriving a regularised reconstruction $\hat{x}$ of the input data $x$, then evaluating both via a scoring function. This function produces an anomaly score $A = f(x, \hat{x})$, which can be thresholded to classify samples as nominal or anomalous or be used directly to inform an engineer about the system’s state. Historically, such scoring functions focused on moment matching; for instance, the squared error $(x - \hat{x})^2$ or the absolute error $\lvert x - \hat{x} \rvert$. These remain popular in recent deep learning approaches, sometimes with rescaling techniques \cite{sarfraz_position_2024,deng_graph_2021} to incorporate homoscedastic uncertainty. However, moment-based methods alone do not fully exploit the probabilistic nature of our VAE framework.

\textbf{Likelihood-based anomaly detection}
Since our training set is assumed to be both anomaly-free and diverse enough to capture the full range of nominal operational conditions, identifying anomalies becomes akin to identifying out-of-distribution samples. A well-trained VAE can learn a calibrated distribution of nominal data. Thus, if a given trace $x$ has a low likelihood under our learned model, it is unlikely to belong to the nominal (in-distribution) regime. Consequently, we measure anomalies using the negative log-likelihood (NLL). Larger NLL values indicate a lower probability of $x$ under the reconstructed distribution, hence a higher degree of anomaly. This follows \citet{an_variational_2015}, who proposed the reconstruction probability, a likelihood-based VAE anomaly score computed from a mean-variance output head. In contrast to their Monte Carlo estimate, which averages the likelihood over multiple latent samples, we evaluate the NLL of a single reconstruction.

In practice, one can determine a suitable threshold on the NLL by examining validation data or specifying acceptable risk levels. This threshold then defines a boundary between nominal and anomalous behaviour, providing a direct, probabilistically grounded metric for fault detection in STS. Additionally, one can use the anomaly scores to manually determine trends and the system's state.

\textbf{Limitations}
A known problem for likelihood-based out-of-distribution detection is that deep generative models can assign higher likelihood to out-of-distribution than to in-distribution inputs \cite{nalisnick2019deep}. This result concerns the marginal model likelihood evaluated on far-from-training distributions, whereas our anomaly score is a reconstruction NLL conditioned on the encoded input, applied to in-domain deviations. The pathology therefore does not transfer directly to our setting.
\subsection{Reconstruction-based Model}
\label{sec:model}
\begin{figure}[pos=htp!]
  \centering
  \includegraphics[width=0.95\linewidth]{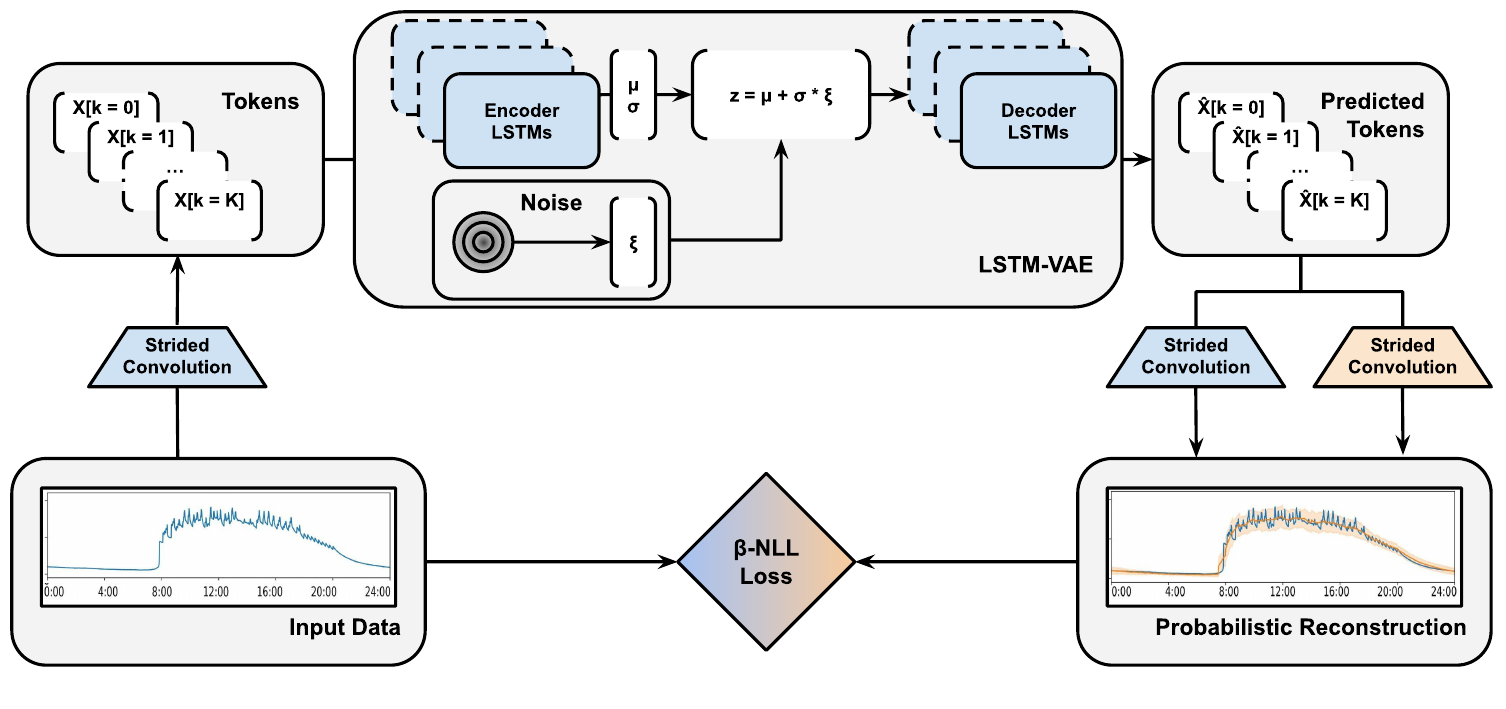}
  \caption{Schematic overview of our LSTM-based VAE for anomaly detection (\texttt{MVE-$\beta$-VAE}). The input time series is tokenised into fixed-length segments and passed through an encoder LSTM, which outputs parameters of a Gaussian latent distribution. A latent sample drawn from this distribution is then fed into a decoder LSTM, which reconstructs the tokens and maps them to a Gaussian distribution, approximating the original input data.}
  \label{fig:model_structure}
\end{figure}

\Cref{fig:model_structure} shows the core idea of an LSTM \cite{hochreiter_long_1997} VAE \cite{kingma_auto-encoding_2014} that reconstructs tokenised sensor data.
Similar work was previously performed on an online anomaly detection task in robotics \cite{park_multimodal_2018}, while other works have used similar but more complex strategies \cite{su_robust_2019}.

\subsubsection{Variational Autoencoder}
\label{sec:variational_autoencoder}
VAEs \cite{kingma_auto-encoding_2014} are probabilistic models widely used in anomaly detection, particularly when the goal is to learn a representation of normal operating conditions and subsequently detect deviations. VAEs extend the standard Autoencoder formulation, which learns a deterministic mapping from data to a bottleneck layer and back, by incorporating a (typically Gaussian) prior distribution over the latent space.

VAEs consist of two main components: an encoder that maps input data $\mathbf{x}$ to a latent representation $\mathbf{z}$, and a decoder that reconstructs $\mathbf{\hat x}$ from samples drawn in the latent space. Formally, the encoder outputs a mean and variance that parameterise a Gaussian distribution over $\mathbf{z}$:
\begin{equation}
  q_\phi(\mathbf{z}|\mathbf{x}) = \mathcal{N}\bigl(\mathbf{z}; \boldsymbol{\mu}_\phi(\mathbf{x}), \boldsymbol{\sigma}_\phi^2(\mathbf{x})\bigr),
\end{equation}
while the decoder models
\begin{equation}
  p_\theta(\mathbf{x}|\mathbf{z}) = \mathcal{N}\bigl(\mathbf{x}; \boldsymbol{\mu}_\theta(\mathbf{z}), \boldsymbol{\sigma}_\theta^2(\mathbf{z})\bigr).
\end{equation}

Training proceeds by minimising a loss that balances reconstruction quality (how accurately the decoder reproduces $\mathbf{x}$) and a Kullback-Leibler (KL) divergence term that encourages $q_\phi(\mathbf{z}|\mathbf{x})$ to remain close to a chosen prior (often $\mathcal{N}(\mathbf{0}, \mathbf{I})$).
An additional hyperparameter $\beta$ is introduced to control the trade-off between reconstruction quality and latent space regularisation in the loss. This results in the $\beta$-VAE loss function \cite{higgins2017betavae}, given by:
\begin{equation}
  \mathcal{L} = -\mathbb{E}_{q_\phi(\mathbf{z}|\mathbf{x})}\left[\log p_\theta(\mathbf{x}|\mathbf{z})\right] + \beta \cdot \mathrm{KL}\left(q_\phi(\mathbf{z}|\mathbf{x}) \| p(\mathbf{z})\right).
\end{equation}
The first term is the expected NLL of the data under the reconstruction and penalises poor reconstruction quality. The second term, $\mathrm{KL}\left(q_\phi(\mathbf{z}|\mathbf{x}) \| p(\mathbf{z})\right)$, measures the divergence between the approximate posterior and the prior distribution. Tuning $\beta$ determines the regularisation strength of the model, with $\beta = 1$ recovering the standard VAE formulation, and a larger $\beta$ parameter leading to stronger restrictions on the latent space. The numerical value of $\beta$ is only meaningful relative to the summation convention of the two loss terms \cite{higgins2017betavae}. We report all $\beta$ values in the standard convention, in which both terms are summed over their respective dimensions.

For a Gaussian output distribution, the NLL per feature and timestep is

\begin{equation}
  -\log p_\theta(x|\mathbf{z}) = \frac{1}{2} \left( \log \sigma^2 + \frac{(x-\mu)^2}{\sigma^2}\right) + \text{const},
\end{equation}

where $\mu$ and $\sigma^2$ are the predicted mean and variance of $\mathbf{\hat x}$ at that position. Under the assumption of constant (homoscedastic) variance, minimising this term is equivalent to the commonly used mean squared error (MSE). Since we are interested in heteroscedastic uncertainty estimation, our model instead predicts both $\mu$ and $\sigma^2$, using a mean-variance estimation (MVE) layer.

As noted by \citet{Seitzer2022Pitfalls}, training with the standard NLL term can lead to issues with undersampling and generally poor training dynamics.
A related problem is reported by \citet{shukla_heteroscedastic_2022} for a VAE with a heteroscedastic output layer, who add a scaled squared-error term to the training objective to avoid poor local optima.
We instead employ the variation of the NLL proposed by \citet{Seitzer2022Pitfalls}, the Beta negative log-likelihood (BNLL), in which each per-element NLL term is weighted by the factor $\lfloor\sigma^{2\beta_L}\rfloor$, where $\lfloor\cdot\rfloor$ denotes the stop-gradient operation. The BNLL is equivalent to the NLL for $\beta_L = 0$, while for $\beta_L = 1$ the mean prediction receives exactly the MSE gradients.

\subsubsection{Data Encoding and Decoding}
Encoding and decoding the time series occur in two main steps, inspired by tokenisation strategies used in computer vision \cite{dosovitskiy2020image}. Since each daily trace contains $1440$ timesteps (one timestep per minute, 1440 minutes per day) and consecutive timesteps often exhibit strong autocorrelation, feeding the entire sequence at once can lead to unstable training dynamics and other problems, resulting in deteriorating performance. Instead, we chunk the data into fixed-length tokens to reduce sequence length while retaining information content.

Specifically, a day’s data consists of $T$ timesteps across $F$ features, which we reshape into $K$ chunks of length $L$ ($K \times L = T$). Each chunk is then mapped by a linear layer into an embedding of size $E$, producing $K$ tokens
\[
  \mathbf{X} \in \mathbb{R}^{T \times F}
  \;\xrightarrow{\text{Reshape}}\;
  \mathbf{X}' \in \mathbb{R}^{K \times L \times F}
  \;\xrightarrow{\text{Linear}}\;
  \mathbf{Z} \in \mathbb{R}^{K \times E}.
\]
Here, each token represents a segment of $L$ timesteps for a given day. The original ordering of tokens is maintained, preserving the temporal structure.

Once the data has been processed into a series of tokens, we use LSTM networks \cite{hochreiter_long_1997} to further process this information and map it to a latent representation. For simplicity of notation, we will now assume that the number of hidden units in the LSTM is equal to the embedding dimension. After processing with the LSTM, the hidden units of the last layer are then linearly mapped to parameterise $D$ Gaussian distributions per token, from which we sample to get a latent vector.
\[
  \begin{aligned}
     & \mathbf{Z} \in \mathbb{R}^{K \times E}
    \;\xrightarrow{\text{LSTM}}\;
    \mathbf{H} \in \mathbb{R}^{K \times E}
    \;\xrightarrow{\text{Linear}}\;
    \boldsymbol{\Theta} \in \mathbb{R}^{K \times D \times 2}%
    \;\xrightarrow{\text{sample}}\;
    \boldsymbol{\Theta}' \in \mathbb{R}^{K \times D}
    .
  \end{aligned}
\]
The first dimension of $\boldsymbol{\Theta}$ represents the mean of each Gaussian distribution, and the second dimension represents its variance.
This retention of a time-dependent latent space is one of the key findings in the OmniAnomaly architecture \cite{su_robust_2019} and also proved beneficial for stable training dynamics.

Decoding reverses the process. The latent vector is linearly mapped to the embedding dimension and processed using the decoder LSTM. A single multi-layer perceptron block (MLP) then produces a mean and standard deviation for each feature and timestep within each token. Finally, we reshape the tokens to restore the original daily format, yielding a probabilistic reconstruction of size $T \times F \times 2$:
\[
  \begin{aligned}
     & \boldsymbol{\Theta}' \in \mathbb{R}^{K \times D}
    \;\xrightarrow{\text{Linear}}\;
    \mathbf{H}' \in \mathbb{R}^{K \times E}
    \;\xrightarrow{\text{LSTM}}\;
    \mathbf{H}'' \in \mathbb{R}^{K \times E}
    \xrightarrow{\text{MLP}}
    \mathbf{R} \in \mathbb{R}^{K \times (L \cdot F) \times 2}
    \;\xrightarrow{\text{reshape}}\;
    \mathbf{\hat{X}} \in \mathbb{R}^{T \times F \times 2}.
  \end{aligned}
\]
Here, the first channel of $\mathbf{\hat{X}}$ corresponds to the reconstructed mean of each feature at each timestep, while the second channel represents the associated variance, providing a probabilistic estimate of the underlying dynamics.
\subsection{Uncertainty Estimation}
\label{sec:uncertainty}
\begin{figure}[pos=htp]
  \centering
  \includegraphics[width=0.7\linewidth]{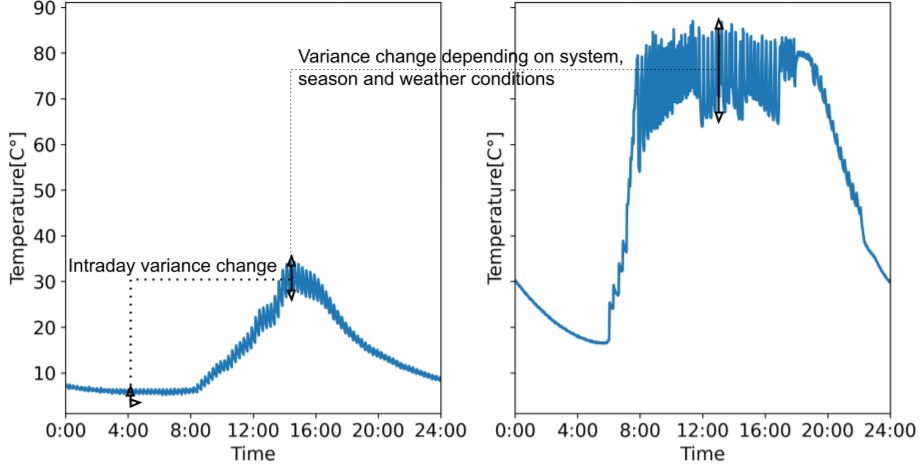}
  \caption{Collector temperature for a high variance (right) and low variance (left) day of the same system. Illustrating the large difference in variance both within a day and between different days.}
  \label{fig:uncertainty}
\end{figure}
Sensor data in STS exhibit variability that depends on both intrinsic (sensor-specific) and extrinsic (state-dependent) factors. For example, the collector's temperature during the day has a broader dynamic range than at night. This variability can further differ based on the system's state and seasonal changes, as shown in \Cref{fig:uncertainty}. Capturing this variability accurately is critical for identifying anomalies, as it enables differentiating between normal fluctuations and genuine faults.

To account for these types of variability, we consider two uncertainty estimation approaches. The first, \textbf{homoscedastic uncertainty estimation}, assumes variability is independent of the data trace and relies solely on the feature-time index. The second, \textbf{heteroscedastic uncertainty estimation}, accounts for data-dependent variability arising from system state, seasonal effects, or other external factors.

\subsubsection{Homoscedastic Uncertainty Estimation}

Homoscedastic uncertainty estimation can be implemented as a post-processing step. Here, the error vector $E_{i,t}$ for sensor $i$ and time step $t$ is calculated using the observed data $\mathbf{x}$ and its reconstruction $\mathbf{\hat{x}}$:
\begin{equation}
  E_{i,t} = \lvert x_{i,t} - \hat{x}_{i,t} \rvert.
\end{equation}

The error vector $E_{i,t}$ can then be normalised using z-normalisation or the inter-quartile transform:
\begin{equation}
  \hat{E}_{i,t} = \frac{E_{i,t} - \mu_{i,t}}{\sigma_{i,t}},
\end{equation}
where $\mu_{i,t}$ is the mean and $\sigma_{i,t}$ is the standard deviation over the training set for z-normalisation, or $\mu_{i,t}$ is the median and $\sigma_{i,t}$ is the inter-quartile range for the inter-quartile transform \cite{sarfraz_position_2024,deng_graph_2021}. This approach assumes a constant, data-independent variance for every feature-time combination,
thereby conforming to the definition of homoscedastic uncertainty. When training a model that gives a deterministic output, we use the standard mean squared error loss.

\subsubsection{Heteroscedastic Uncertainty Estimation}
While post-processing methods provide a simple way to estimate uncertainty, they fail to capture data-dependent variability. For example, \Cref{fig:uncertainty} demonstrates that the dynamic range of collector temperature varies substantially across different system states and seasons. To address this, we incorporate uncertainty estimation into the model by using an MVE layer.
In this approach, the model predicts both the mean and variance of a Gaussian distribution for each feature and timestep from the input data, thereby learning heteroscedastic (data-dependent) variance.

An MVE-based approach can adapt to evolving conditions, such as seasonal shifts or changing operational states, by providing a probabilistic reconstruction at each timestep. This is particularly relevant for domestic STS, where day-to-day behaviour can vary widely depending on ambient conditions, system state, and usage patterns. We employ a mean-variance approach for simplicity, allowing the model to output a Gaussian parameterised by $(\mu, \sigma)$ per timestep and feature, without requiring any additional hyperparameters. When training a model outputting a Gaussian distribution, it is necessary to choose an appropriate loss function. As discussed in \Cref{sec:variational_autoencoder}, we choose the BNLL here for improved training dynamics. Since no gradients are involved at evaluation time, these considerations do not apply there, and we score anomalies with the standard NLL, which directly corresponds to the model likelihood. An ablation of the $\beta_L$ parameter, covering the special cases of training with the NLL ($\beta_L = 0$) and the MSE ($\beta_L = 1$), is provided in \Cref{sec:appendix_betal}.

\section{Experiments}
\subsection{Dataset}
\label{sec:dataset}
We used the PaSTS dataset \cite{ebmeier_pasts_2024}, consisting of $83$ different STS with varying installations and operational durations, for a total of 39878 days (corresponding to approx. 110 system-years).
The data is taken from the manufacturer's,  Ritter Energie GmbH \& Co. KG,  service team and thus contains many anomalies. Only a few systems are considered faultless or mostly faultless under normal operational conditions.
Four data types are provided: index, sensor, control signals, and status. The index data specify the system and timestep, sensor data records direct readings, and control signals and status data document the controller’s behaviour, including outputs from an internal expert system. All measurements are at a one-minute resolution. For a more detailed summary, see \Cref{sec:appendix_dataset}.

\textbf{Dataset reannotation}
The dataset contains data from multiple types of installations, various old and new controllers, fault detection algorithms, and system types. Certain systems exhibit multi-year gaps in data collection or have undergone renovations and modifications over time.
Because this dataset primarily reflects service operations, many annotations from the built-in fault detection system are either incomplete or inaccurate, as properly identified and resolved faults do not require additional analysis by the service team and are handled by technicians or system operators. Consequently, we treated these annotations as unreliable.

During early stages of the project we found many unannotated faults in the dataset, which led us to reannotate parts of it in collaboration with two domain engineers from the manufacturer. To direct expert attention, days were screened using early versions of \texttt{MVE-$\beta$-VAE} as well as the \texttt{Rescaled PCA-R} model. The final judgement was made by the engineers based on the raw sensor readings alone. The reannotation was strictly one-directional, recasting days initially labelled as fault-free into fault conditions only where the engineers identified clear, yet initially unrecognised, anomalies. Ambiguous cases were left unchanged. Due to the size of the dataset, not all systems were revisited, and false annotations remain. However, the most problematic examples, which severely distorted the summary statistics, have been corrected.
In total, 2170 days across 29 systems were relabelled. The five most affected systems account for 1927 of these days ($\approx 89\%$), corresponding to long-running faults that had gone undetected by the built-in expert system. \Cref{tab:reannotation_audit} summarises the relabelled days per system.

\textbf{Dataset split along systems}
We separated our data into traces of a single day, with a total of $1440$ minutes of data per trace.
From an operational perspective, we assume that a fault detection method should be able to handle unseen systems from their first day of data onwards. We separated training, validation, and test data in our dataset along different systems, such that the entire evaluation is always done on unseen data traces from unseen systems.
In our experiments we only trained on nominal data, thereby constraining our training set to contain systems in a nominal operational state, with all potential anomalies and faults filtered out.
We then separated the dataset into training, validation and test set, with the training and validation set only containing mostly faultless systems and the test set containing systems with both nominal and faulty behaviour. The split into these different sets was always done along systems, such that a system is always fully contained in only one of the three sets. The splits can be found in \Cref{sec:appendix_splits}.

\textbf{Limitations}
The PaSTS dataset contains no fault-type annotations. Detection performance can therefore not be stratified by fault type, and complementary strengths between methods, with different methods detecting different fault types, remain invisible in the aggregate metrics.
Similarly, the system types are not recorded and the sensor set is homogeneous across systems, so generalisation to different system realisations and sensor availabilities cannot be assessed. Future work would benefit from a dataset with extensive metadata on fault types and system realisations.

\subsection{Baselines}
\label{sec:experiments_baselines}
We implemented several baselines to compare our work to. We used the PCA reconstruction method as demonstrated by \citet{sarfraz_position_2024}, an LSTM-based VAE utilising homoscedastic uncertainty estimation as well as several baselines as implemented in the recent benchmark publication by \citet{NEURIPS2024_c3f3c690}.

\textbf{PCA Reconstruction}
\citet{sarfraz_position_2024} have shown that PCA can outperform the state-of-the-art deep learning models, while being very easy to tune and train. Analogous to their implementation, we fitted the principal components on the training set. Then we calculated the error vector between the data and the reconstructed data using the fitted PCA and fitted a z-transformation of this vector, as described earlier.
On the test set, we transformed the data according to the fitted PCA model, reconstructed it to calculate the error vector, and then rescaled the error vector using the fitted z-transformation. The average of the error vector then yields a single anomaly score per day. Later, we refer to this method as \texttt{Rescaled PCA-R} and the method without z-transforming the error vector as \texttt{Unscaled PCA-R}.
The PCA reconstruction method involves only a single hyperparameter, namely the number of principal components used for reconstruction.
To compare the ease of use, we swept this hyperparameter and evaluated the difficulty of optimisation. We then compare it to the parameter sweep of $\beta$ of the main model and discuss the optimisation difficulty for the main model.

\textbf{Benchmark baselines}
As additional baselines, we utilised the implemented baselines from the recent benchmark publication by \citet{NEURIPS2024_c3f3c690}. As part of their work, they implement several multivariate anomaly detection methods. We used the hyperparameters and setup they provide in the GitHub repository accompanying their publication. We then integrated the results into our problem definition, analysing the aggregated anomaly score for an entire day. The implemented methods are several statistical methods, as well as neural network-based methods, based on density, reconstruction, or forecasting. The two forecasting-based methods (\texttt{CNN} and \texttt{LSTMAD}) are comparatively simple and, according to the findings by \citet{sarfraz_position_2024}, we would expect them to perform well.
The \texttt{PCA} method implemented in the paper is not a reconstruction-based method but a density-based method within principal component space.
Furthermore, we updated the TranAD \cite{tuli2022tranad} implementation provided in the GitHub repository to include correct normalisation. We also removed the OmniAnomaly \cite{su_robust_2019} implementation, which lacked essential components of the original implementation.
\subsection{Experimental Setup}
\label{sec:experimental_setup}
We trained all models using only nominal STS, as identified in \Cref{sec:appendix_splits}, and validated them on a separate nominal dataset. The input variables include five temperature sensors (\texttt{TSA1}, \texttt{TSE}, \texttt{TW}, \texttt{TSV}, \texttt{TAM}), a volumetric flowrate sensor (\texttt{VF}), and two control signals (\texttt{pwm} for pump operation and \texttt{ctr} for frost protection). We $z$-normalised the temperature sensors based on training statistics, and rescaled the remaining sensors to $(0,1)$, smoothing them with a 15-minute Gaussian window to stabilise training. \citet{ebmeier_pasts_2024} provide further details on the data distribution.

We chose a set of practical hyperparameters, as summarised in \Cref{tab:hyperparams}, and trained the model on a single NVIDIA A100 GPU. The primary hyperparameter sweep concerned the latent space regularisation factor $\beta$, covering $\beta \in [0, 6]$, from the unregularised case ($\beta = 0$) to values above the standard VAE ($\beta = 1$).\footnote{In our implementation, the reconstruction term is averaged over the $L \cdot F$ elements of each token while the KL term is summed over the latent dimensions. The loss coefficient $\alpha$ in the released code therefore relates to the reported $\beta$ as $\beta = \alpha \cdot L \cdot F$.}
At convergence, the KL term on the validation set is $4.61$\,nats per token (${\approx}0.58$\,nats per latent dimension; see \Cref{sec:appendix_training} for the training curves), confirming that the posterior has not collapsed and the latent regularisation is active.
We report an additional sweep of the BNLL parameter ($\beta_L$) in \Cref{sec:appendix_betal}.

We evaluated the model quantitatively using four different metrics. We used the standard optimal F1 score, AUC-PR and AUC-ROC, as they are commonly used in time series anomaly detection (e.g. \cite{sarfraz_position_2024,NEURIPS2024_c3f3c690}). Additionally, we calculated the system-wise F1 score by taking the threshold leading to the optimal F1 score for all systems except the one we looked at and then applying this threshold to the individual system. We then calculated the F1 score per system and averaged the F1 score over all systems. In the following, we refer to this as the system-wise F1 score. This metric mirrors our application most closely, as it quantifies the detection performance of a new unseen system being evaluated. For a short discussion of the metrics, see \Cref{sec:appendix_metrics}. We trained all stochastic models with $N = 10$ different seeds and report the mean and standard deviation. The deterministic methods have no seed variance, and we report them as point values. Figures and tables showing a single training run of \texttt{MVE-$\beta$-VAE} use the median seed with respect to the system-wise F1 score. For each metric, we tested every model against the best-performing model, using Welch's two-sided t-test between two stochastic models and a one-sample t-test against the point value of a deterministic method, at a significance level of $0.05$.

While the system-wise F1 score mirrors the transfer of a threshold to an unseen system, its threshold selection still uses labelled test data. In \Cref{sec:appendix_percentile}, we additionally report results under fixed thresholds calibrated only on the nominal validation set.

\begin{table}[ht]
  \centering
  \caption{Hyperparameters used in \texttt{MVE-$\beta$-VAE}.}
  \label{tab:hyperparams}
  \begin{tabular}{l@{\quad}c}
    \toprule
    \textbf{Hyperparameter}             & \textbf{Value} \\
    \midrule
    Latent Space Regularisation $\beta$ & 2.4            \\
    BNLL Parameter $\beta_L$            & 0.5            \\
    Learning Rate                       & 1e-4           \\
    LSTM Layers                         & 4              \\
    Hidden Dimension                    & 64             \\
    Latent Dimension                    & 8              \\
    Dropout                             & 0.1            \\
    Token Length                        & 30             \\
    Update Steps                        & 40\,000        \\
    \bottomrule
  \end{tabular}
\end{table}

Our evaluation focuses on three main hypotheses.
\begin{enumerate}
  \item Reconstruction-based anomaly detection can adequately detect faults in STS.
  \item Our method can match or exceed the anomaly detection performance of both simple baselines, such as PCA, and more complex established DL methods. This would be surprising given the findings of \citet{sarfraz_position_2024}.
  \item Uncertainty estimation improves anomaly detection performance. Here, we evaluate whether heteroscedastic uncertainty estimation is superior to homoscedastic uncertainty estimation.
\end{enumerate}

\subsection{Quantitative Results}
\label{sec:quantitative_results}
\Cref{tab:results1} compares the performance of \texttt{MVE-$\beta$-VAE} against several baselines on the reannotated dataset, which
incorporates expert feedback from the manufacturer. \Cref{tab:results2} reports the same comparison on the original dataset annotations. \Cref{sec:appendix_reconstructions} shows some example reconstructions by \texttt{MVE-$\beta$-VAE}. Besides listing each method’s optimal F1 score and the previously defined system-wise F1 score, we report the area under the precision-recall curve (AUC-PR) and the area under the receiver operating characteristic (AUC-ROC).

\begin{table*}[ht]
  \centering
  \caption{Comparison of anomaly detection performance on the reannotated dataset. Values are the mean over all seeds ± the standard deviation. For non-stochastic methods---like PCA---no standard deviation is given. Baselines are as discussed in \Cref{sec:experiments_baselines}. Baselines marked with $\dagger$ are forecasting-based deep learning models and forecast a single step. The best-performing model per metric and all models without a statistically significant difference to it are marked in bold (see \Cref{sec:experimental_setup}). Pairs of deterministic methods cannot be tested under this procedure.}
  \label{tab:results1}
  \begin{tabular}{l l@{\,}r|l@{\,}r|l@{\,}r|l@{\,}r}
    \toprule
    \textbf{Method}                                   & \multicolumn{2}{@{}c|}{System-wise F1} & \multicolumn{2}{c|}{Optimal F1} & \multicolumn{2}{c|}{AUC-PR} & \multicolumn{2}{c}{AUC-ROC}                                                                   \\
    \midrule
    \texttt{Unscaled PCA-R}                           & 0.30                                   &                                 & 0.69                        &                             & 0.74          &          & 0.86             &                   \\
    \texttt{Rescaled PCA-R}                           & 0.40                                   &                                 & 0.76                        &                             & \textbf{0.82} &          & \textbf{0.91}    &                   \\
    \texttt{MSE-$\beta$-VAE}                          & 0.32 ±                                 & 0.02                            & 0.72  ±                     & 0.01                        & 0.75  ±       & 0.01     & 0.89  ±          & $<$ 0.01          \\
    \texttt{MVE-$\beta$-VAE}                          & \textbf{0.46} ±                        & \textbf{0.01}                   & \textbf{0.78} ±             & \textbf{0.01}               & 0.80  ±       & 0.01     & \textbf{0.90}  ± & \textbf{$<$ 0.01} \\
    \midrule
    \texttt{HBOS\cite{goldstein2012histogram}}        & 0.15                                   &                                 & 0.50                        &                             & 0.52          &          & 0.74             &                   \\
    \texttt{LOF\cite{breunig2000lof}}                 & 0.29                                   &                                 & 0.48                        &                             & 0.47          &          & 0.77             &                   \\
    \texttt{KMeansAD\cite{yairi2001fault}}            & 0.24 ±                                 & 0.02                            & 0.57 ±                      & 0.01                        & 0.62  ±       & 0.02     & 0.82  ±          & 0.01              \\
    \texttt{PCA\cite{paffenroth2018robust}}           & 0.38                                   &                                 & 0.75                        &                             & 0.79          &          & 0.88             &                   \\
    \texttt{IForest\cite{liu2008isolation}}           & 0.15                                   &                                 & 0.57                        &                             & 0.60          &          & 0.75             &                   \\
    \midrule
    \texttt{AnomalyTransformer\cite{xu_anomaly_2022}} & 0.28 ±                                 & 0.02                            & 0.35 ±                      & 0.03                        & 0.32 ±        & 0.06     & 0.62 ±           & 0.08              \\
    \texttt{AutoEncoder\cite{sakurada2014anomaly}}    & 0.24 ±                                 & $<$ 0.01                        & 0.30 ±                      & $<$ 0.01                    & 0.22 ±        & $<$ 0.01 & 0.57 ±           & $<$ 0.01          \\
    \texttt{CNN\cite{munir2018deepant}}$\dagger$      & 0.36 ±                                 & 0.01                            & 0.75 ±                      & 0.01                        & 0.78 ±        & 0.01     & 0.88 ±           & $<$ 0.01          \\
    \texttt{Donut\cite{tran2016distance}}             & 0.30 ±                                 & 0.03                            & 0.69 ±                      & 0.01                        & 0.71 ±        & 0.01     & 0.84 ±           & 0.01              \\
    \texttt{LSTMAD\cite{malhotra2015long}}$\dagger$   & 0.30 ±                                 & 0.01                            & 0.65 ±                      & 0.02                        & 0.71 ±        & 0.02     & 0.84 ±           & 0.01              \\
    \texttt{TimesNet\cite{wu2022timesnet}}            & 0.25 ±                                 & 0.01                            & 0.36 ±                      & 0.01                        & 0.38 ±        & 0.01     & 0.56 ±           & 0.01              \\
    \texttt{TranAD\cite{tuli2022tranad}}              & 0.32 ±                                 & 0.02                            & 0.68 ±                      & 0.03                        & 0.72 ±        & 0.03     & 0.85 ±           & 0.01              \\
    \texttt{USAD\cite{audibert_usad_2020}}            & 0.25 ±                                 & 0.02                            & 0.37 ±                      & 0.02                        & 0.29 ±        & 0.01     & 0.67 ±           & 0.01              \\
    \bottomrule
  \end{tabular}
\end{table*}

The \texttt{Unscaled PCA-R} approach serves as a simple baseline, achieving a system-wise F1 score of $0.30$ and an optimal F1 score of $0.69$. Rescaling the error vector (\texttt{Rescaled PCA-R}) leads to a substantial increase in performance over all metrics, outperforming all other baselines that we compare to on the reannotated dataset. This indicates that even using homoscedastic uncertainty estimation greatly improves performance and aligns with previous studies, which have also observed a clear performance improvement when employing this method \cite{sarfraz_position_2024,deng_graph_2021}.

Transitioning from PCA to a mean-only VAE (\texttt{MSE-$\beta$-VAE}) yields an improved performance level over the \texttt{Unscaled PCA-R} baseline. However, even though this method also uses the error-rescaling, it does not improve performance substantially over the unscaled method and falls behind \texttt{Rescaled PCA-R}.

When comparing to \texttt{MVE-$\beta$-VAE}, a more substantial gain becomes apparent. As the model also estimates variance, we computed the anomaly score using the NLL. \texttt{MVE-$\beta$-VAE} achieves the best optimal F1 score of 0.78, 0.02 ahead of \texttt{Rescaled PCA-R}, while \texttt{Rescaled PCA-R} remains slightly ahead on AUC-PR. The difference on AUC-ROC is not statistically significant. The largest difference is in the application-focused system-wise F1 score, where \texttt{MVE-$\beta$-VAE} outperforms all other models by a clear margin. This metric is the strongest indicator of generalisation between systems and thus mirrors our generalisation-focused approach. As an additional analysis, \Cref{tab:confusion} in \Cref{sec:appendix_metrics} shows the confusion matrices of \texttt{MVE-$\beta$-VAE} and \texttt{Rescaled PCA-R} at the optimal threshold.
Furthermore, \texttt{MVE-$\beta$-VAE} outperforms all the baselines we adapted from \citet{NEURIPS2024_c3f3c690} on all metrics.

\begin{table*}[ht]
  \centering
  \caption{Comparison of anomaly detection performance on the original dataset annotation. Values are the mean over all seeds ± the standard deviation. For non-stochastic methods---like PCA---no standard deviation is given. Baselines are as discussed in
    \Cref{sec:experiments_baselines}. Baselines marked with $\dagger$ are forecasting-based deep learning models and forecast a single step. The best-performing model per metric and all models without a statistically significant difference to it are marked in bold (see \Cref{sec:experimental_setup}). Pairs of deterministic methods cannot be tested under this procedure.}
  \label{tab:results2}
  \begin{tabular}{l l@{\,}r|l@{\,}r|l@{\,}r|l@{\,}r}
    \toprule
    \textbf{Method}                                   & \multicolumn{2}{@{}c|}{System-wise F1 } & \multicolumn{2}{c|}{Optimal F1} & \multicolumn{2}{c|}{AUC-PR} & \multicolumn{2}{c}{AUC-ROC}                                                                             \\
    \midrule
    \texttt{Unscaled PCA-R}                           & 0.26                                    &                                 & 0.40                        &                             & 0.38            &                   & 0.77            &                   \\
    \texttt{Rescaled PCA-R}                           & \textbf{0.28}                           &                                 & 0.40                        &                             & 0.28            &                   & 0.78            &                   \\
    \texttt{MSE-$\beta$-VAE}                          & 0.24 ±                                  & 0.03                            & 0.37 ±                      & 0.01                        & 0.36 ±          & 0.02              & \textbf{0.79} ± & \textbf{$<$ 0.01} \\
    \texttt{MVE-$\beta$-VAE}                          & \textbf{0.30} ±                         & \textbf{0.03}                   & \textbf{0.43} ±             & \textbf{0.03}               & 0.33 ±          & 0.04              & \textbf{0.79} ± & \textbf{0.01}     \\
    \midrule
    \texttt{HBOS\cite{goldstein2012histogram}}        & 0.09                                    &                                 & 0.20                        &                             & 0.11            &                   & 0.59            &                   \\
    \texttt{LOF\cite{breunig2000lof}}                 & 0.16                                    &                                 & 0.34                        &                             & 0.29            &                   & 0.75            &                   \\
    \texttt{KMeansAD\cite{yairi2001fault}}            & 0.18 ±                                  & 0.01                            & 0.33 ±                      & $<$ 0.01                    & 0.30 ±          & 0.01              & 0.76 ±          & $<$ 0.01          \\
    \texttt{PCA\cite{paffenroth2018robust}}           & 0.25                                    &                                 & 0.35                        &                             & 0.18            &                   & 0.73            &                   \\
    \texttt{IForest\cite{liu2008isolation}}           & 0.07                                    &                                 & 0.20                        &                             & 0.15            &                   & 0.60            &                   \\
    \midrule
    \texttt{AnomalyTransformer\cite{xu_anomaly_2022}} & 0.19 ±                                  & 0.02                            & 0.28 ±                      & 0.04                        & 0.22 ±          & 0.04              & 0.67 ±          & 0.04              \\
    \texttt{AutoEncoder\cite{sakurada2014anomaly}}    & 0.16 ±                                  & $<$ 0.01                        & 0.20 ±                      & $<$ 0.01                    & 0.16 ±          & $<$ 0.01          & 0.59 ±          & $<$ 0.01          \\
    \texttt{CNN\cite{munir2018deepant}}$\dagger$      & 0.25 ±                                  & 0.01                            & \textbf{0.44} ±             & \textbf{0.01}               & \textbf{0.39} ± & \textbf{$<$ 0.01} & \textbf{0.79} ± & \textbf{$<$ 0.01} \\
    \texttt{Donut\cite{tran2016distance}}             & 0.17 ±                                  & 0.02                            & 0.36 ±                      & 0.02                        & 0.31 ±          & 0.01              & 0.71 ±          & 0.01              \\
    \texttt{LSTMAD\cite{malhotra2015long}}$\dagger$   & 0.23 ±                                  & 0.02                            & 0.41 ±                      & $<$ 0.01                    & \textbf{0.40} ± & \textbf{0.01}     & 0.78 ±          & $<$ 0.01          \\
    \texttt{TimesNet\cite{wu2022timesnet}}            & 0.18 ±                                  & 0.02                            & 0.36 ±                      & 0.01                        & 0.35 ±          & 0.01              & 0.70 ±          & 0.01              \\
    \texttt{TranAD\cite{tuli2022tranad}}              & 0.17 ±                                  & 0.02                            & 0.38 ±                      & 0.02                        & 0.38 ±          & 0.02              & 0.76 ±          & 0.01              \\
    \texttt{USAD\cite{audibert_usad_2020}}            & 0.18 ±                                  & 0.01                            & 0.23 ±                      & $<$ 0.01                    & 0.22 ±          & $<$ 0.01          & 0.64 ±          & 0.01              \\
    \bottomrule
  \end{tabular}\end{table*}

\Cref{tab:results2} shows the results for the original dataset annotation. We observe the same general behaviour as on the reannotated dataset: \texttt{MVE-$\beta$-VAE} achieves the best system-wise F1 and an optimal F1 score statistically indistinguishable from the best.
We performed the reannotation together with industry professionals and consider it more reliable than the original annotation.
The forecasting-based models perform better on the original annotation, with the \texttt{CNN} matching \texttt{MVE-$\beta$-VAE} on the optimal F1 score and exceeding it on AUC-PR. The disagreement between the two annotations is heavily concentrated. The five most relabelled systems account for 89\% of the relabelled days (\Cref{tab:reannotation_audit}) and the metrics under the original annotation are correspondingly affected by these systems for all models alike.

\begin{figure}[pos=h]
  \centering
  \includegraphics[width=0.6\linewidth]{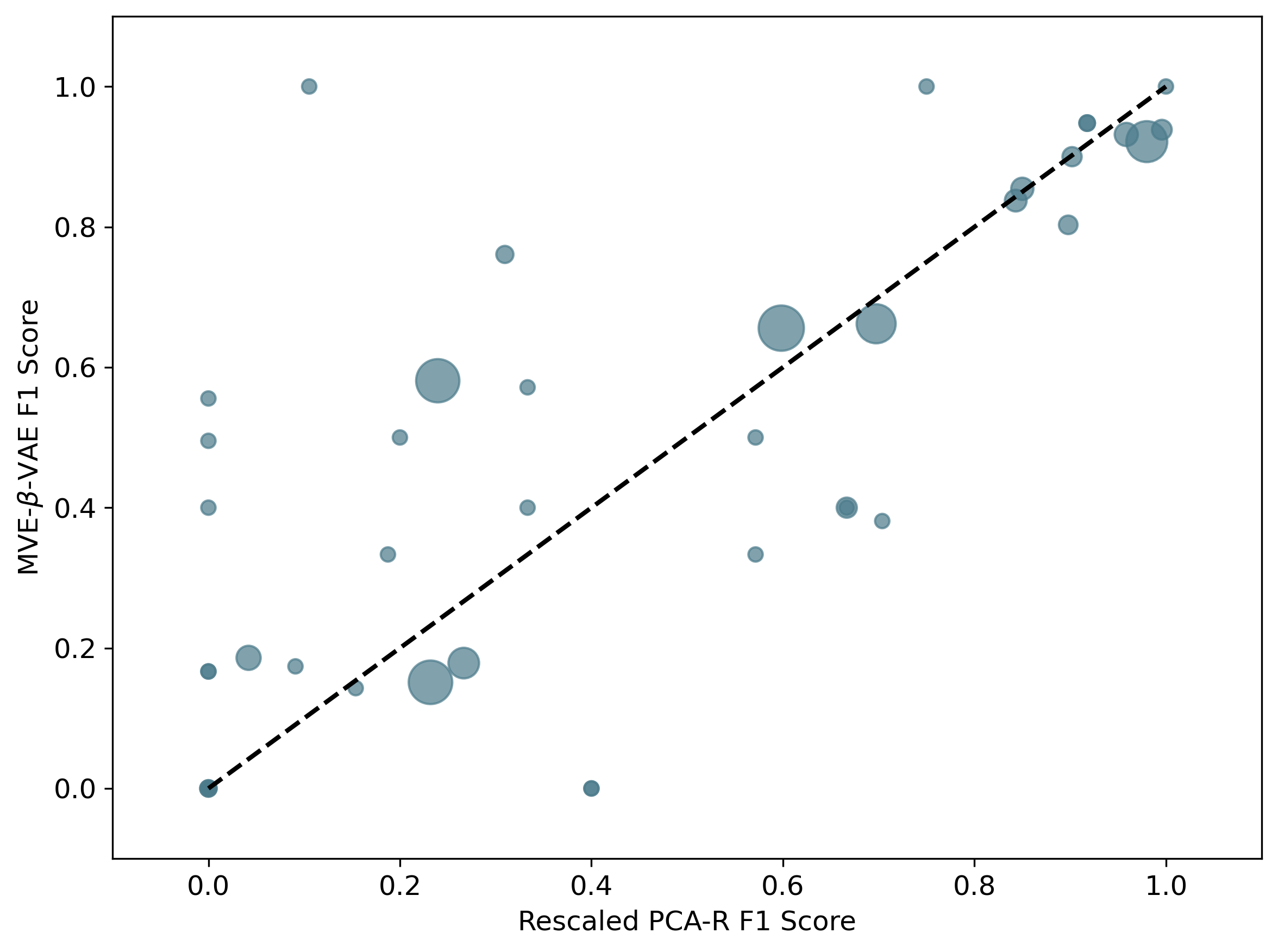}
  \caption{Per-system F1 score of \texttt{MVE-$\beta$-VAE} versus \texttt{Rescaled PCA-R} on the reannotated dataset. Each marker is one system. The marker area is proportional to the number of days of data for that system. Points above the dashed diagonal indicate systems on which \texttt{MVE-$\beta$-VAE} achieves the higher F1 score.}
  \label{fig:system_wise_comparison}
\end{figure}

In \Cref{fig:system_wise_comparison}, we compare the per-system F1 scores of \texttt{MVE-$\beta$-VAE} and \texttt{Rescaled PCA-R}. Each marker corresponds to one test system, with the marker area proportional to the number of days recorded for that system. The system-wise F1 score is the average along the respective axis. Most systems lie close to the diagonal, where both models perform comparably. The largest deviations lie above the diagonal, including several systems on which \texttt{Rescaled PCA-R} fails almost entirely while \texttt{MVE-$\beta$-VAE} detects a substantial share of the faults. Deviations in favour of \texttt{Rescaled PCA-R} are less frequent but not negligible, reaching differences of up to $0.40$ in per-system F1. For a more detailed comparison between \texttt{MVE-$\beta$-VAE} and \texttt{Rescaled PCA-R} as well as an additional qualitative analysis, see \Cref{Sec:PCA}.

Overall, these results are in good agreement with our key hypotheses. Reconstruction-based anomaly detection achieves solid quantitative results on our dataset. \texttt{MVE-$\beta$-VAE} achieves the best system-wise F1 score among all evaluated methods, which include reconstruction-based, forecasting-based, and density-based models. Both heteroscedastic and homoscedastic uncertainty estimation substantially enhance detection performance. While the models perform comparably on many systems, there are systems where \texttt{MVE-$\beta$-VAE} strongly outperforms all baselines. We show this in the comparison to \texttt{Rescaled PCA-R} in \Cref{Sec:PCA}.

The performance of \texttt{MVE-$\beta$-VAE} compared to the \texttt{MSE-$\beta$-VAE} also demonstrates the benefit of heteroscedastic uncertainty estimation for reconstruction-based anomaly detection.
Since \texttt{MVE-$\beta$-VAE} differs from the \texttt{MSE-$\beta$-VAE} in both the training loss and the anomaly score, we decoupled the two contributions in an additional ablation (\Cref{sec:appendix_betal}, \Cref{tab:score_ablation}).
Scoring with the NLL instead of the MSE accounts for the larger share of the gain in system-wise F1, while training with the BNLL adds a further improvement in both metrics. The gain from NLL scoring is concentrated in the system-wise F1, reinforcing that heteroscedastic uncertainty estimation is central to generalisation across systems. The predicted variances are well-calibrated on held-out nominal data (\Cref{sec:appendix_calibration}), with a spread-skill ratio of $0.98$.
A caveat of heteroscedastic NLL scoring is that, by construction, deviations of the mean are down-weighted in regimes where the model predicts high variance, so faults occurring in such regimes could in principle be masked. We did not observe an instance of this in our evaluation, but as not all days were manually revisited (\Cref{sec:dataset}), we cannot rule it out.

\subsection{Qualitative Results}
\label{sec:qualitative_results}
\begin{figure}[pos=htp!]
  \centering
  \begin{subfigure}[b]{0.49\linewidth}
    \centering
    \includegraphics[width=\linewidth]{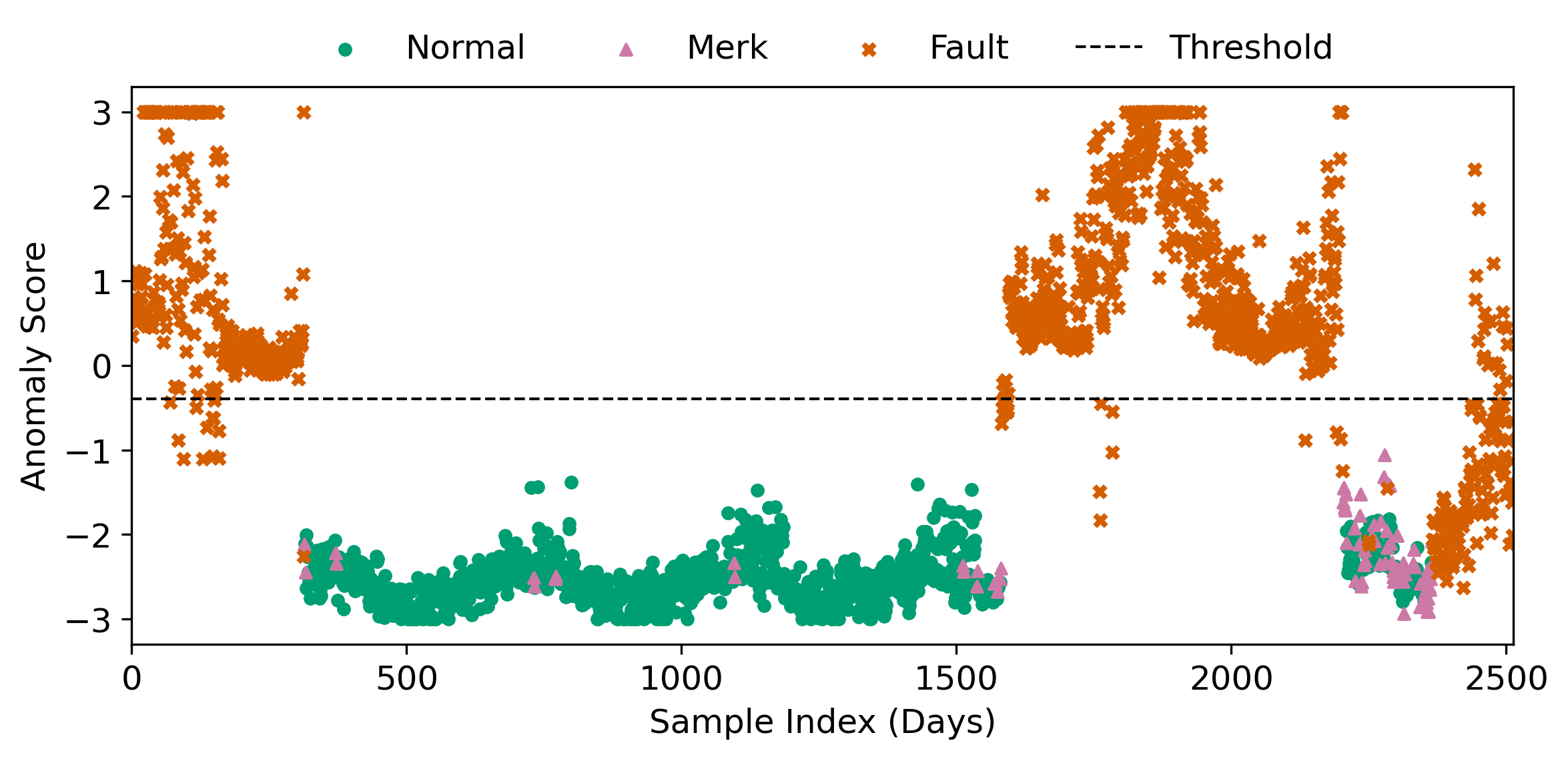}
    \caption{System~40 (F1 0.92) exhibits a long, previously undetected fault, leading to reduced yield and eventual degradation of multiple components.}
    \label{fig:qual1}
  \end{subfigure}
  \hfill
  \begin{subfigure}[b]{0.49\linewidth}
    \centering
    \includegraphics[width=\linewidth]{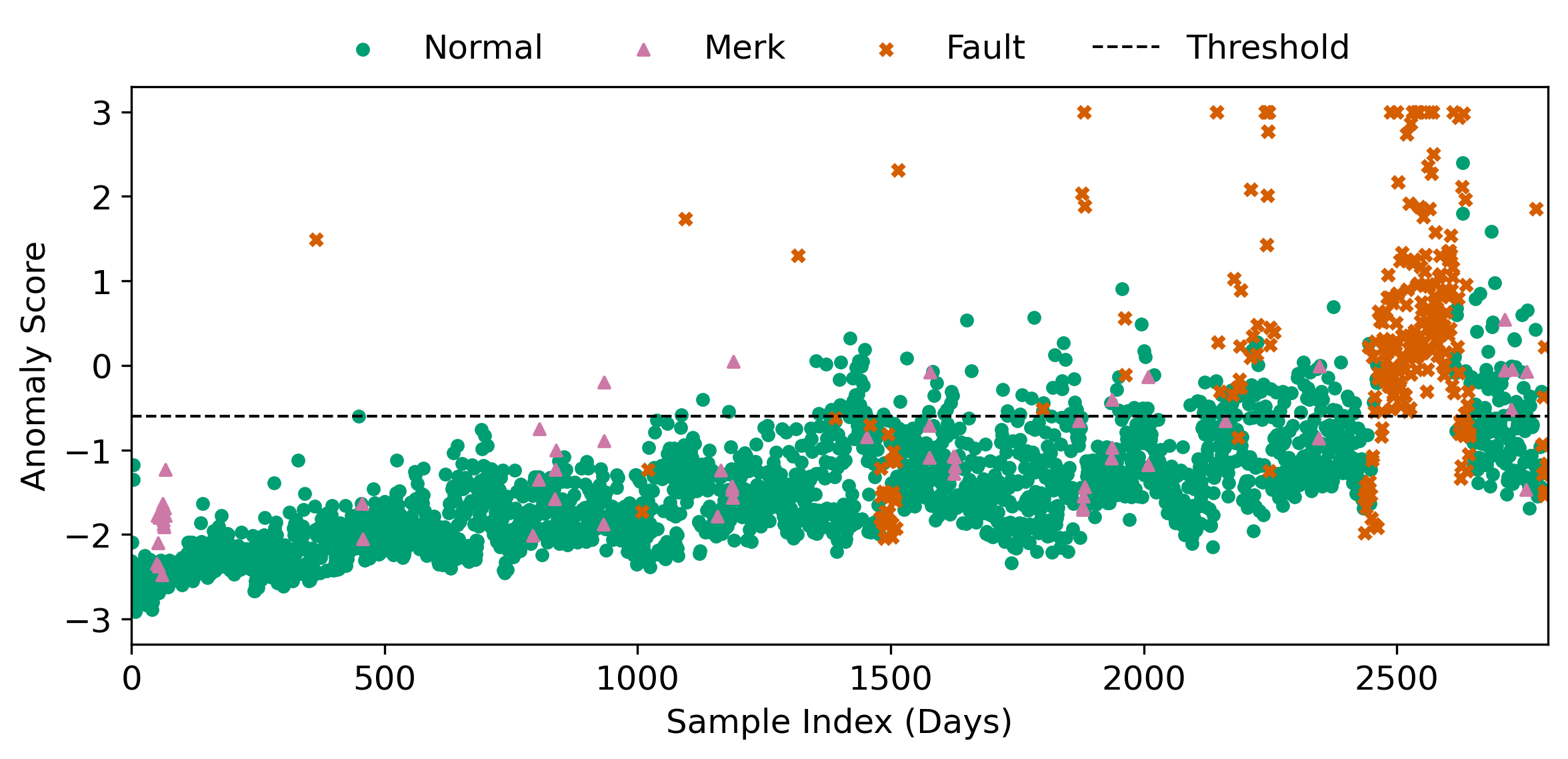}
    \caption{System~52 (F1 0.58) shows no clear boundary between fault and nominal states, though the anomaly score indicates a run-to-failure trend near the end of its life cycle.}
    \label{fig:qual2}
  \end{subfigure}
  \caption{Qualitative results for two sample systems. The x-axis represents the day index, and the y-axis is the anomaly score (values above 3 are capped for visibility). Individual dots denote a day with a fault annotation (Fault), a potential anomaly marked by the expert system (Merk) or nominal behaviour (Normal) according to its colour.}
  \label{fig:qual}
\end{figure}
To validate our quantitative findings, we investigated two example systems in the dataset that are known to alternate between nominal and faulty behaviour.
In \Cref{fig:qual1}, we show System~40 from the dataset. The x-axis represents the index of the days, while the y-axis displays the anomaly score, capped at $3$ for clearer visualisation. Days with a fault annotation, with a potential anomaly, or with nominal behaviour are marked with their respective colours in the figure. A clear separation emerges between nominal and faulty behaviour. After a manual analysis of this system, in collaboration with the manufacturer, we discovered a previously undetected, long-persisting fault that lasted over 600 days (around day indices 1\,529 to 2\,130). This fault further degraded the system, leading to multiple faults towards the end of the time series. \texttt{MVE-$\beta$-VAE} thus contributed to uncovering these hidden issues and provided valuable insights into the system's state.

An additional example can be seen in \Cref{fig:qual2}, illustrating System~52. Here, the separation between nominal and faulty states is less pronounced. Nonetheless, the system exhibits a run-to-failure pattern, with more anomalous or faulty days towards the end of its life cycle. A qualitative inspection of the anomaly score time series still helps reveal the gradual degradation into an anomalous state, even if automated detection does not yield perfect results for all days. Hence, visual inspection of the anomaly score remains a valuable tool for monitoring the system’s status and has proven informative in our analysis of the dataset.

Faults that push signals outside their nominal operational range, such as the sensor lock in \Cref{fig:fault_day}, produce exceptionally high anomaly scores. In contrast, some fault states lie inside the nominal data distribution itself. For example, a fault-induced shutdown resembles a dormant state that also occurs during nominal operation. Such faults are reconstructed accurately and are detectable only through contextual inconsistencies.

In \Cref{Sec:PCA}, we compare the qualitative performance of \texttt{MVE-$\beta$-VAE} with \texttt{Rescaled PCA-R}. Notably, for the above-mentioned System~52, the anomaly score of \texttt{MVE-$\beta$-VAE} exhibits a rising trend well before the system failure, enabling early detection, while no such trend is present in \texttt{Rescaled PCA-R} (see \Cref{fig:comparison_52}). Additionally, we observe that \texttt{Rescaled PCA-R} fails on some systems, while \texttt{MVE-$\beta$-VAE} achieves excellent performance (\Cref{fig:comparison_14}: F1 0.50 vs.\ 0.00; \Cref{fig:comparison_107}: 1.00 vs.\ 0.11; \Cref{fig:comparison_119}: 0.56 vs.\ 0.00). This mirrors the performance of \texttt{MVE-$\beta$-VAE} in the system-wise F1 metric. Given the usefulness of the anomaly score itself, \texttt{MVE-$\beta$-VAE} is currently being implemented into the manufacturer's service team workflows.

\subsection{Optimisation Complexity}
\label{sec:optimisation}
\begin{figure}
  \centering
  \begin{subfigure}[b]{0.49\linewidth}
    \centering
    \includegraphics[width=\linewidth]{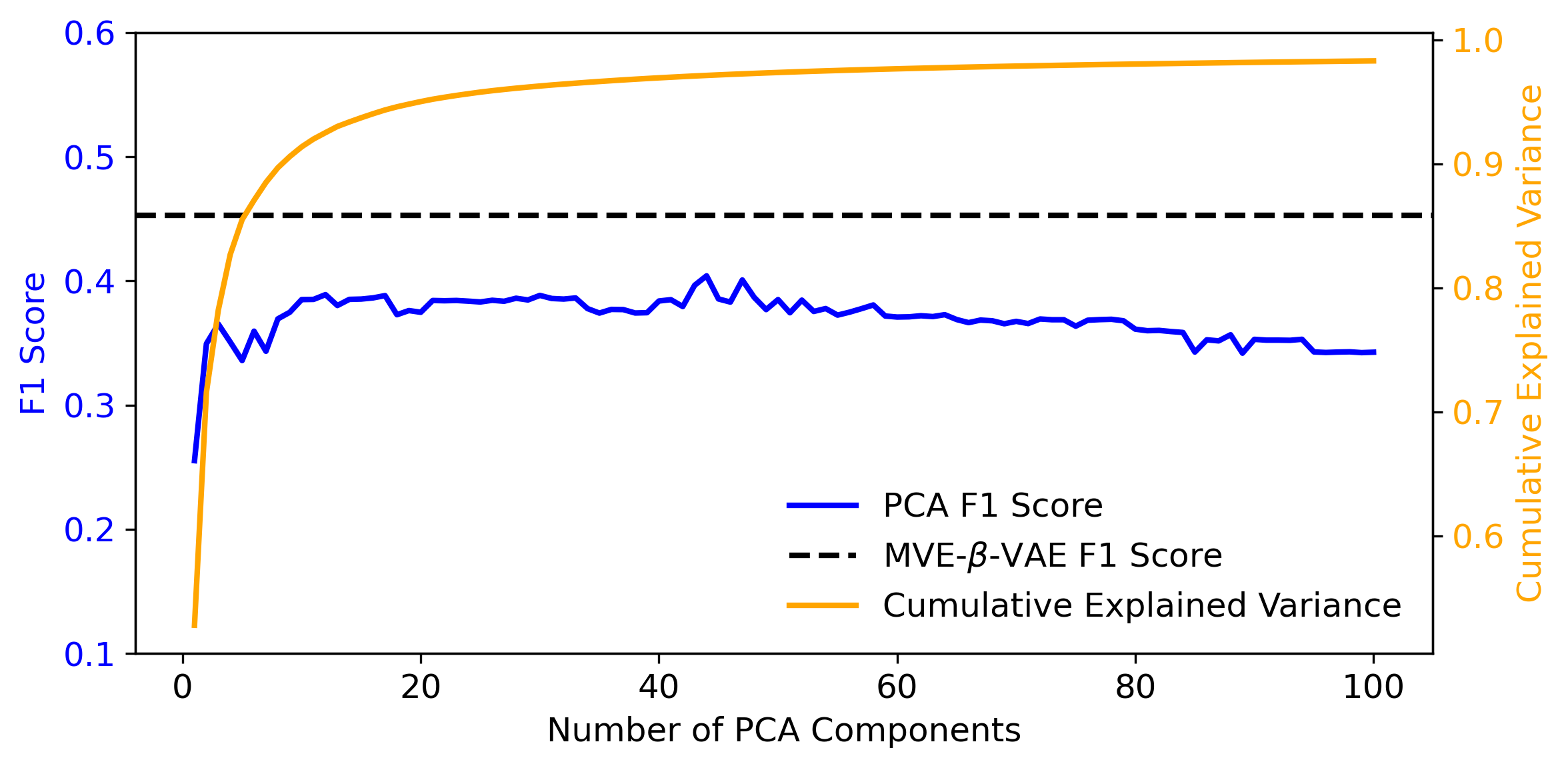}
    \caption{System-wise F1 optimisation.}
  \end{subfigure}
  \hfill
  \begin{subfigure}[b]{0.49\linewidth}
    \centering
    \includegraphics[width=\linewidth]{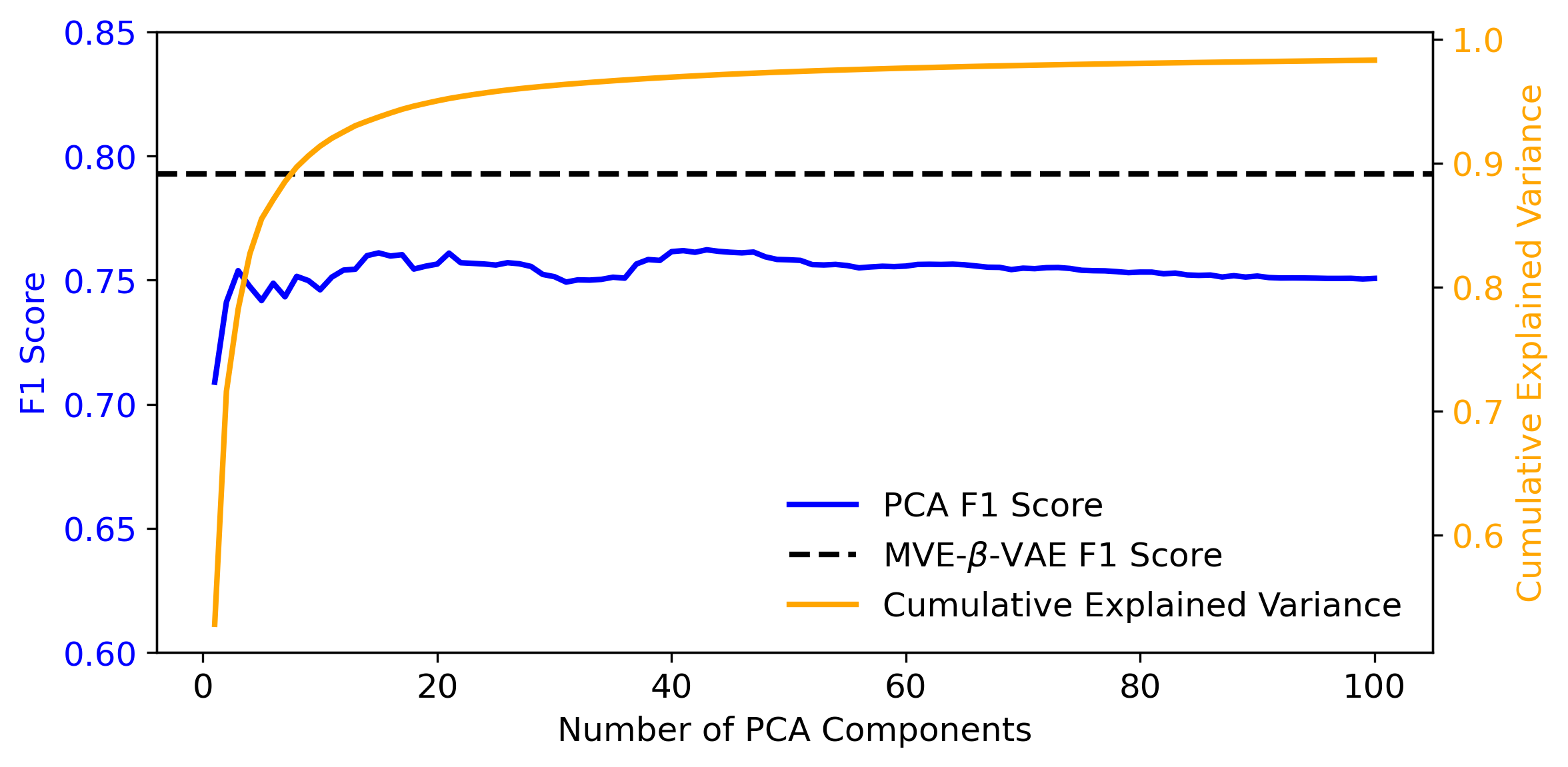}
    \caption{Optimal F1 optimisation.}
  \end{subfigure}
  \caption{Impact of the number of principal components on anomaly detection performance for the \texttt{Rescaled PCA-R} model. The x-axis depicts the number of components, and the y-axis shows both the resulting F1 score and the cumulative variance explained by those components.}
  \label{fig:pca_optim}
\end{figure}

\Cref{fig:pca_optim} illustrates how the single hyperparameter in the \texttt{Rescaled PCA-R} baseline, the number of principal components, affects anomaly detection performance. The x-axis shows the number of components, while the y-axis tracks both the F1 score and the cumulative variance explained. Identifying the near-optimal range requires under 30 minutes of runtime, even with unoptimised code. As shown, using only three principal components already yields an optimal F1 score of 0.75, with the maximum at 0.76. This simple parameter sweep underscores how straightforward it is to tune PCA in practice, making it unlikely to miss the optimal setting.

\begin{figure}
  \centering
  \begin{subfigure}[b]{0.49\linewidth}
    \centering
    \includegraphics[width=\linewidth]{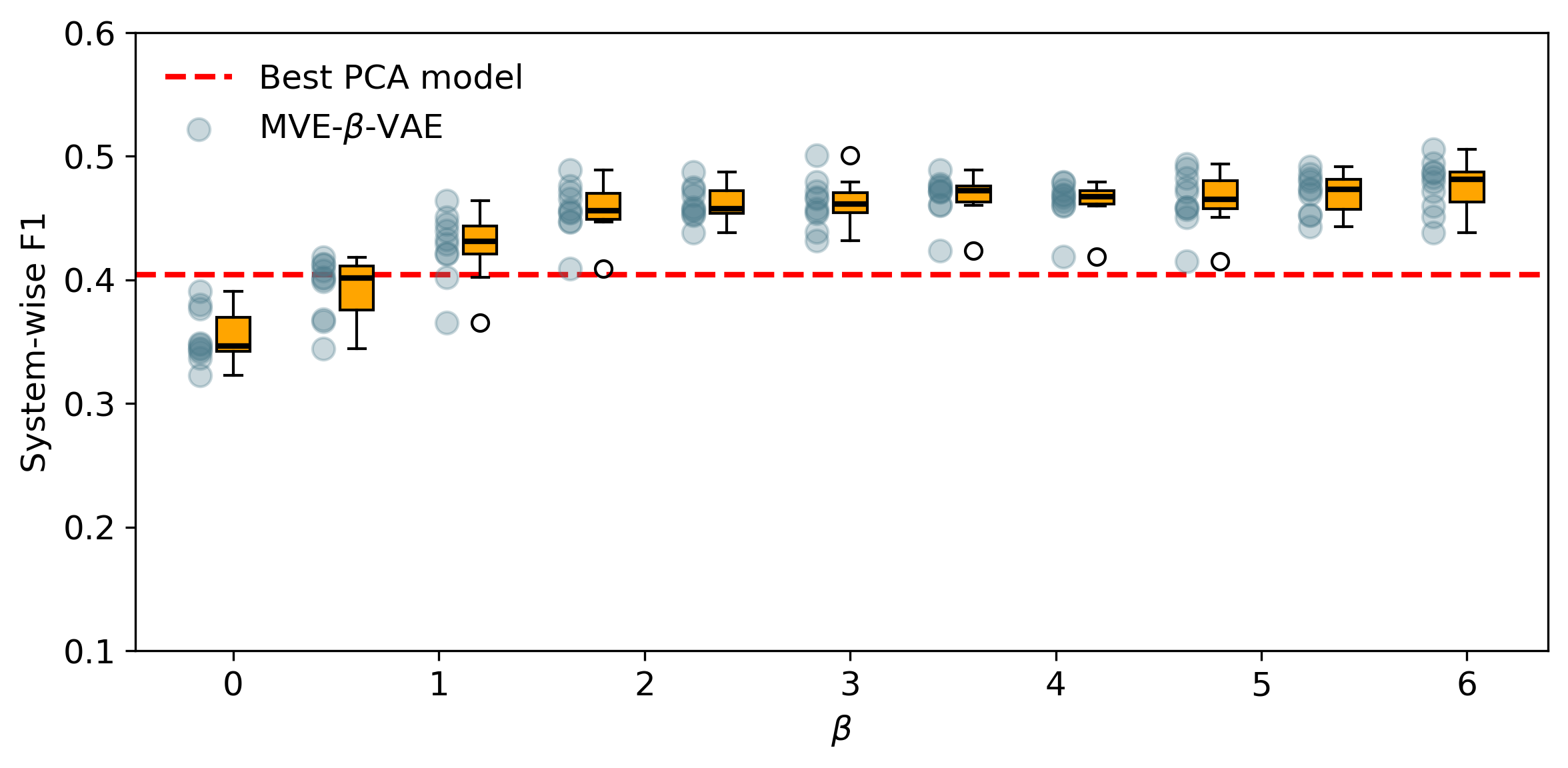}
    \caption{System-wise F1 optimisation.}
  \end{subfigure}
  \hfill
  \begin{subfigure}[b]{0.49\linewidth}
    \centering
    \includegraphics[width=\linewidth]{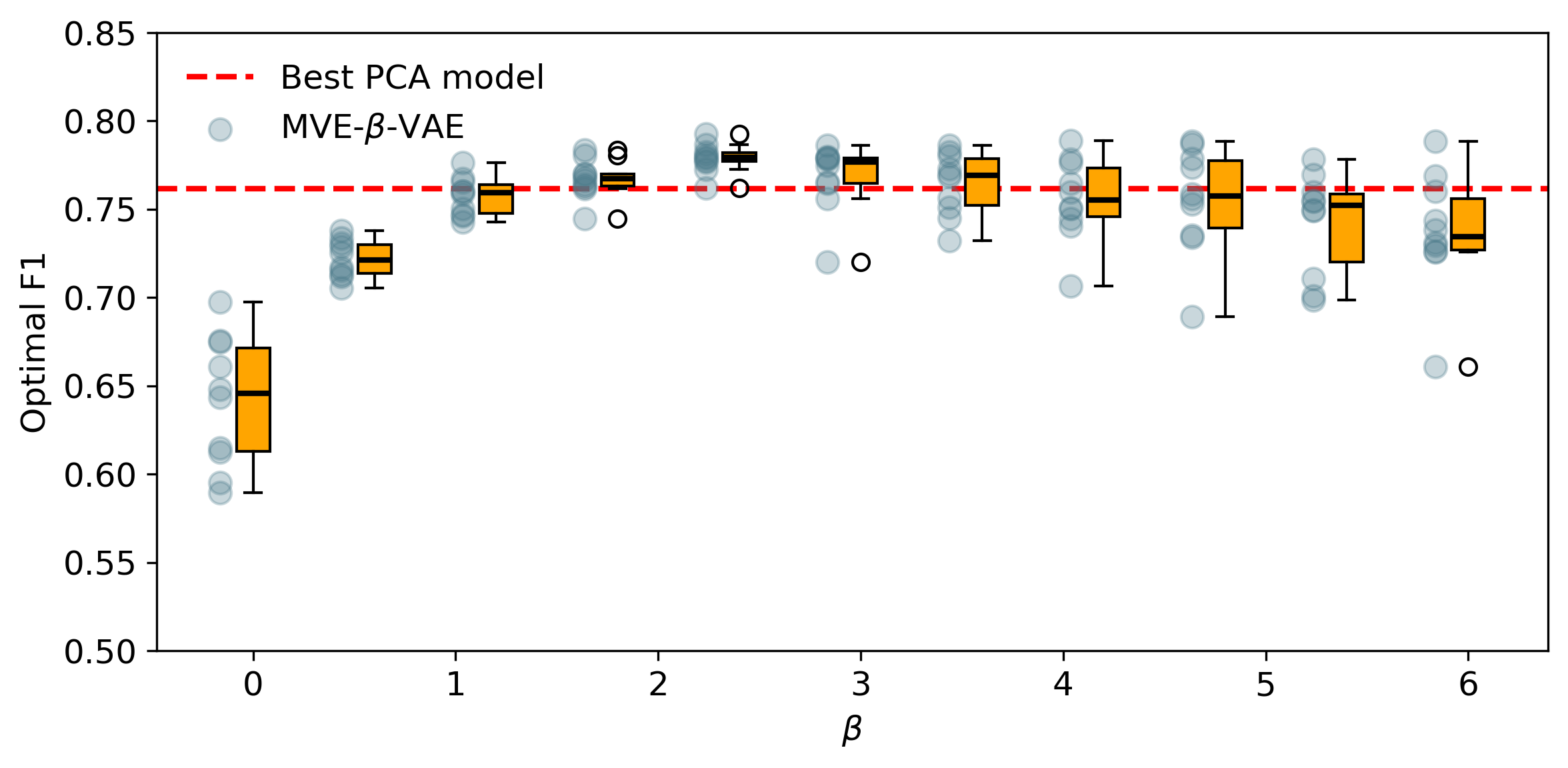}
    \caption{Optimal F1 optimisation.}
  \end{subfigure}
  \caption{Impact of the latent space regularisation parameter $\beta$ on the F1 score. Each dot represents a single training run, highlighting how both hyperparameter choices and initialisation affect model performance. The box extends from the first quartile to the third quartile. The line inside the box corresponds to the median. The whiskers extend from the box to the largest datapoint within 1.5 inter-quartile ranges from the box. The flier points are the points beyond the ends of the whiskers. The dashed red line marks the best \texttt{Rescaled PCA-R} baseline. $\beta = 0$ corresponds to a stochastic Autoencoder with a Gaussian output head, $\beta = 1$ to a standard VAE.}
  \label{fig:model_optim}
\end{figure}

By contrast, our VAE-based approach introduces multiple hyperparameters and more intricate engineering requirements. \Cref{fig:model_optim} illustrates how just one of these hyperparameters, the latent space regularisation parameter $\beta$, influences the F1 score. Each dot corresponds to a single seed training run, while the dashed red line marks the best \texttt{Rescaled PCA-R} baseline. Even for the same $\beta$ value, the range of observed F1 scores is considerably larger than the variation in \texttt{Rescaled PCA-R}. In many cases, training a VAE multiple times with the same hyperparameter configuration is necessary to obtain consistent results.
Additionally, the position of the shown optimum can be affected by other hyperparameters such as the size of the latent space, making optimisation over multiple parameters even more compute- and time-intensive.

Stable training for long time series relies on careful weight initialisation, normalisation layers, and other deep learning techniques that may not be immediately familiar to engineers without machine learning expertise. In contrast, the \texttt{Rescaled PCA-R} baseline requires none of these specialised considerations or high-end hardware. While deep learning models can surpass simple baselines, increasing model complexity increases both the engineering overhead and computational cost, emphasising the trade-off between potential performance gains and ease of development.

\section{Conclusion}
\label{sec:conclusion}
In this work, we pose three key questions: Can reconstruction-based anomaly detection effectively identify faults in STS? Can VAEs match or surpass simple and complex baselines in time series anomaly detection? Can uncertainty estimation boost fault detection performance? Our qualitative findings show distinct separations between nominal and faulty states, even when faults persist for hundreds of days. These observations are further confirmed by the manufacturer's experts, who indicate that a day-by-day reconstruction-based workflow can reliably uncover anomalies in real operational data. This is also the reason that the model is currently being implemented into the service workflows of the manufacturer.

Quantitative results reinforce this view, demonstrating that \texttt{MVE-$\beta$-VAE} achieves the best system-wise F1 score among all evaluated methods, with a substantial portion of this performance attributed to explicitly modelling heteroscedastic uncertainty. In this application-focused metric, which quantifies the transfer to unseen systems, the gains over all baselines are largest, highlighting the generalisation capabilities of \texttt{MVE-$\beta$-VAE}. This is a surprising result given the findings by \citet{sarfraz_position_2024} and highlights that deep learning, if properly configured, can outperform simpler linear approaches on generalisation. Nonetheless, \texttt{Rescaled PCA-R} remains a robust baseline that trails \texttt{MVE-$\beta$-VAE} by only 0.02 on optimal F1 and stays slightly ahead on AUC-PR with minimal tuning.

Our evaluation shows the benefits of modelling heteroscedastic uncertainty within a VAE. The ablation of training loss and scoring function shows that the heteroscedastic variance estimates drive the gain in cross-system generalisation, with NLL scoring accounting for the larger share. These probabilistic reconstructions do not increase the overall model complexity.

At the same time, deep learning architectures demand considerable engineering effort. Multiple hyperparameters can critically affect performance. Finding an optimal configuration may require extensive parameter sweeps and retraining, which can be prohibitive in practice. In contrast, simple linear methods like PCA often involve only a few key hyperparameters, offering a more straightforward route to acceptable results with minimal domain knowledge.

Consequently, we make a case for simple models to increase the performance and usability of anomaly detection models. This holds particularly for model choices such as the heteroscedastic uncertainty estimation via MVE, which adds no additional hyperparameters. While deep learning can achieve higher performance, the gain comes at the cost of greater engineering overhead and computational expense. Many advanced models in time series anomaly detection, featuring novel architectures and arbitrary parameterisations, might be too complex to feasibly reach their parametric optimum. As a result, they may underperform against baseline or “simplified” methods---an outcome consistent with the results by \citet{sarfraz_position_2024} and \citet{NEURIPS2024_c3f3c690}.
Thus, elaborate network architectures that increase complexity can be a hindrance that should be avoided during development. Nevertheless, deep learning methods for anomaly detection also enable other time series analysis tasks and may be further extended into various downstream tasks.

In summary, these findings confirm the feasibility of using probabilistic reconstruction-based anomaly detection for STS. Future research should emphasise approaches that balance performance gains with ease of development, ensuring that hyperparameter optimisation does not become prohibitively complex and that new methods can be assessed on a level playing field against other methods.

\section*{Data availability}
The data used for this study is available under \url{https://zenodo.org/records/11093493}, see also \cite{ebmeier_pasts_2024}. The code and the reannotation file will be made available upon publication.

\section*{Declaration of competing interest}
The authors declare that they have no known competing financial interests or personal relationships that could have influenced the work reported in this paper.

\section*{Declaration of generative AI and AI-assisted technologies in the manuscript preparation process}
During the preparation of this work the authors used LLMs (like ChatGPT, Grammarly) for writing assistance, including grammar correction and text polishing. After using this tool/service, the author(s) reviewed and edited the content as needed and take(s) full responsibility for the content of the published article.

\section*{Acknowledgements}
This project was funded by the Deutsche Forschungsgemeinschaft (DFG, German Research Foundation) under Germany's Excellence Strategy - EXC number 2064/1 - Project number 390727645. The authors thank the international Max Planck Research School for Intelligent Systems (IMPRS-IS) for supporting Florian Ebmeier.
We want to thank Ritter Energie for providing us with the data on which this dataset is based. We additionally thank Ritter Energie for their expertise in reannotating the dataset and evaluating our models.

\printcredits
\bibliographystyle{elsarticle-harv}
\bibliography{references}

\par\ifdim\dimexpr\pagegoal-\pagetotal\relax<12\baselineskip \newpage\fi %
\appendix
\crefalias{section}{appendix} %
\renewcommand\thefigure{A\arabic{figure}}
\renewcommand\thetable{A\arabic{table}}
\setcounter{figure}{0}
\setcounter{table}{0}

\section*{Appendix}
\titlecontents{lsection}[2em]{\small}{\contentslabel{2em}}{}{\titlerule*[0.6pc]{.}\contentspage}
\startcontents[appendices]
\printcontents[appendices]{l}{1}{\setcounter{tocdepth}{1}}
\newpage

\section{Train, Validation, and Test Splits}
\label{sec:appendix_splits}

Our dataset is divided into three disjoint sets: training, validation, and testing. Each set comprises entire systems, ensuring that no system appears in more than one set. This setup reflects our goal of testing fault detection on systems unseen during training.

\textbf{Training Systems}

\noindent
\texttt{[001, 013, 023, 025, 033, 035, 041, 043, 049, 053, 058, 060, 062, 069, 075, 077, 105, 106, 110, 116, 126, 138, 146, 150, 173, 175, 193, 902, 904, 905]}

\textbf{Validation Systems}

\noindent
\texttt{[026, 038, 046, 063, 074, 076, 114, 118, 136]}

\textbf{Test Systems}

\noindent
\texttt{[002, 003, 005, 007, 012, 014, 015, 016, 027, 028, 029, 030, 031, 034, 036, 037, 040, 044, 045, 047, 048, 050, 051, 052, 054, 059, 061, 064, 065, 067, 068, 070, 072, 107, 119, 144, 149, 154, 158, 166, 169, 179, 181, 903]}

This partition ensures that each subset contains entire systems, preventing data leakage across training, validation, and test phases. By testing on systems that were never seen during training, we simulate a real-world scenario in which the model encounters previously unknown systems.

\section{PaSTS Dataset}
\label{sec:appendix_dataset}
In this work, we used the PaSTS dataset \cite{ebmeier_pasts_2024}. The dataset is taken from the manufacturer's service team and consists of systems in a completely nominal state, systems degrading from a nominal state towards a fault state, and systems only in fault states. The dataset consists of $83$ different STS with varying installations and operational durations between $1$ and $3329$ days of operation, for a total of 39878 days.
We used the preprocessed dataset, which includes only full days, with a time resolution of $1$ datapoint per minute, for a total of $1440$ datapoints per day. For every timestep, there are a total of 17 fields. These fields are index data (the timestamp and the system identifier), sensor data (temperature, yield and volumetric flow data), control signals to the pump, and status indicators.
Which of the variables we used in our experiments is discussed in \Cref{sec:experimental_setup}.

The status indicators include the output of an internal expert system, which is a rules- and physics-based system, that classifies the current system state into nominal, anomalous (\texttt{Merk}) or a fault (\texttt{sto}). The controller propagates anomalous and fault states over time and automatically runs tests that confirm or reject these states. Additionally, for some classified states, the system will shut itself down, leading to passive, nominal behaviour.

\Cref{tab:reannotation_audit} contains the relabelled days per system resulting from the reannotation described in \Cref{sec:dataset}. The five most affected systems account for 1927 of the 2170 relabelled days, corresponding to long-running faults that had gone undetected by the built-in expert system.

\begin{table}[pos=h!]
  \centering
  \caption{Relabelled days per system during the dataset reannotation ($2170$ days in total). Only systems with at least one relabelled day are listed; these are the systems in which the
    engineers identified clear, previously unannotated faults. Systems without relabelled days were either not revisited, contained only ambiguous cases, or were already sufficiently well annotated.}
  \label{tab:reannotation_audit}
  \begin{tabular}{cc@{\qquad}cc@{\qquad}cc@{\qquad}cc}
    \toprule
    \textbf{System} & \textbf{Days} & \textbf{System} & \textbf{Days} & \textbf{System} & \textbf{Days} & \textbf{System} & \textbf{Days} \\
    \midrule
    2               & 10            & 31              & 10            & 50              & 5             & 68              & 149           \\
    3               & 10            & 36              & 10            & 51              & 6             & 149             & 2             \\
    5               & 149           & 37              & 405           & 52              & 182           & 158             & 1             \\
    7               & 34            & 40              & 1042          & 59              & 3             & 179             & 3             \\
    15              & 12            & 44              & 1             & 61              & 7             & 903             & 2             \\
    27              & 7             & 45              & 1             & 64              & 3             &                 &               \\
    29              & 66            & 47              & 4             & 65              & 1             &                 &               \\
    30              & 25            & 48              & 6             & 67              & 14            &                 &               \\
    \bottomrule\end{tabular}
\end{table}

\FloatBarrier
\newpage
\section{Example Reconstructions}
\label{sec:appendix_reconstructions}
Here we give some example reconstructions of a system in a nominal state and a system in a faulty state using \texttt{MVE-$\beta$-VAE}. The reconstruction quality can be tuned to be arbitrarily good. The reconstruction quality is only interesting for us in the context of anomaly detection, and reconstructions should only be directly used to analyse the behaviour of the model, looking for unexpected behaviour.
\begin{figure}[pos=htp!]
  \centering
  \includegraphics[width=0.7\linewidth]{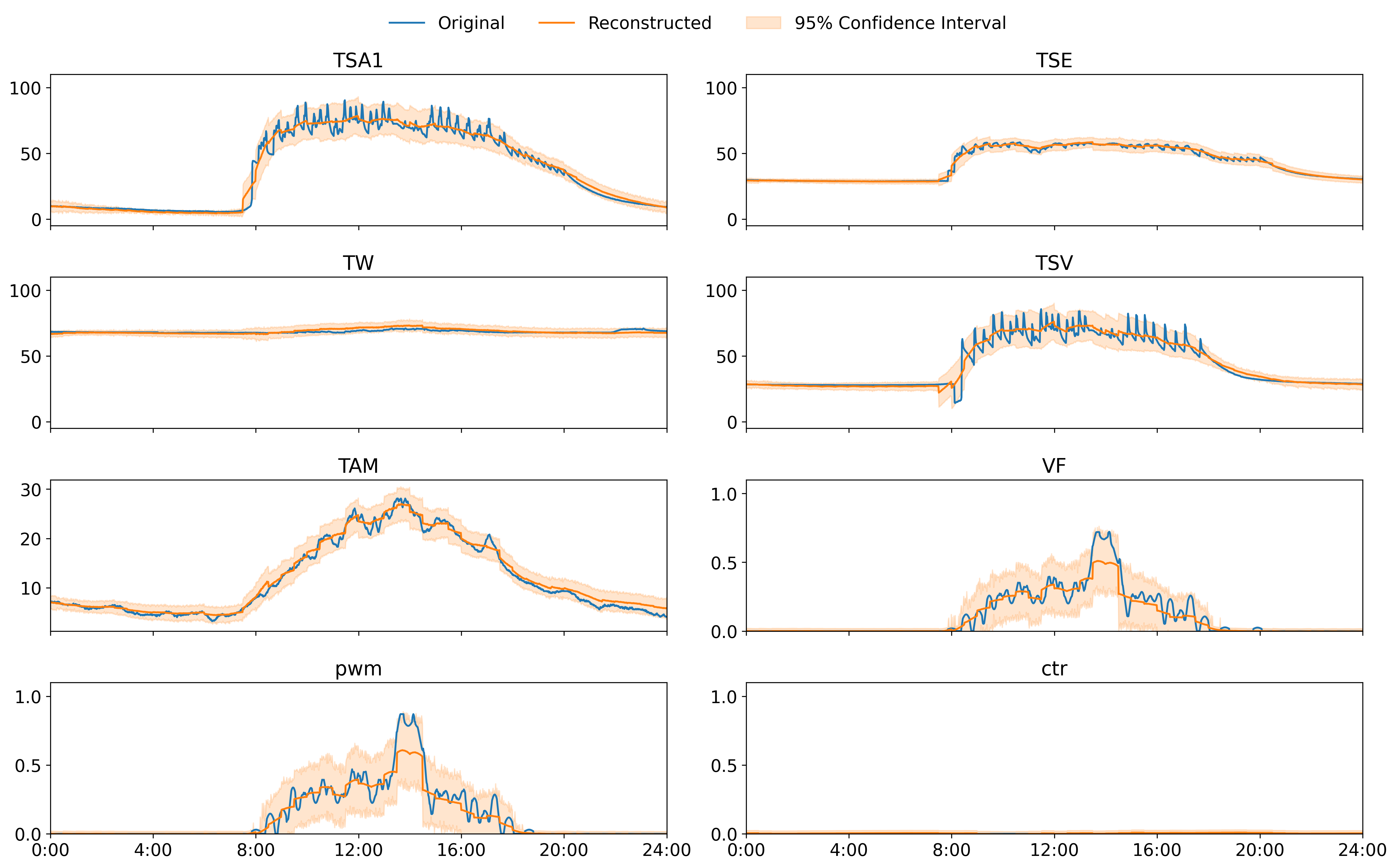}
  \caption{Example of the reconstruction of a normal spring day (System~40, 23 March 2017, anomaly score $-2.56$). Here the system is working exactly as intended, which results in good reconstructions and low anomaly scores.}
  \label{fig:normal_day}

  \vspace{\baselineskip}
  \includegraphics[width=0.7\linewidth]{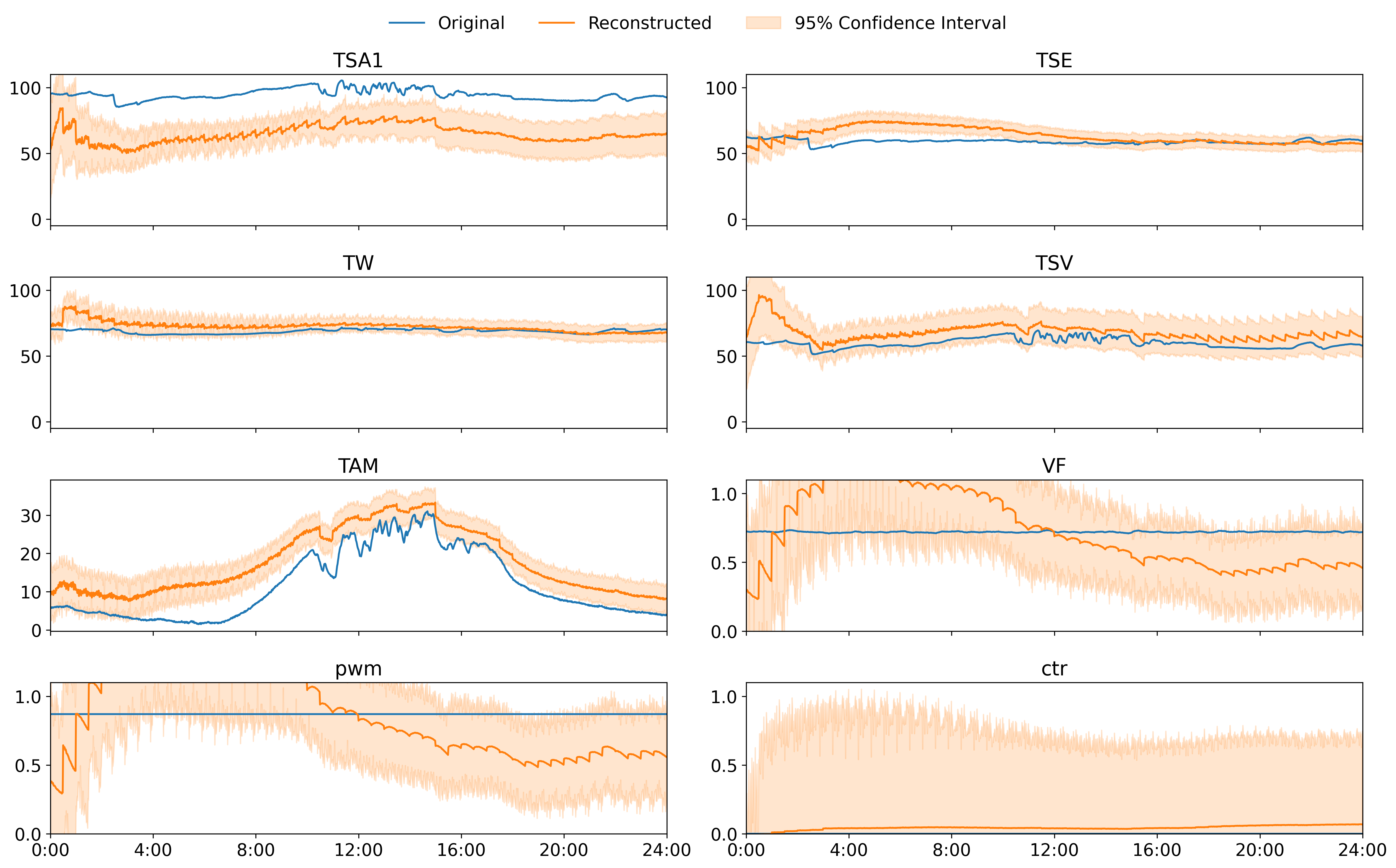}
  \caption{Example of the reconstruction of a faulty day (System~40, 25 September 2018, anomaly score $16.75$). Due to an electrical issue, the system entered a state in which the \texttt{TSA1} sensor is consistently reading the wrong temperature. This leads to a locked state, in which the pump remains permanently running, resulting in increased heating demand and more severe faults down the line. As this is completely out of distribution, the reconstructions are off, and we observe high anomaly scores.}
  \label{fig:fault_day}
\end{figure}

\FloatBarrier
\newpage

\section{Comparison of \texttt{Rescaled PCA-R} and \texttt{MVE-$\beta$-VAE}}
\label{Sec:PCA}
Comparison of the best performing baseline---\texttt{Rescaled PCA-R} model---with \texttt{MVE-$\beta$-VAE}.
\begin{figure}[pos=htp!]
  \centering
  \begin{subfigure}[b]{0.49\linewidth}
    \centering
    \includegraphics[width=\linewidth]{system_52_mve.png}
    \caption{\texttt{MVE-$\beta$-VAE}.}
  \end{subfigure}
  \hfill
  \begin{subfigure}[b]{0.49\linewidth}
    \centering
    \includegraphics[width=\linewidth]{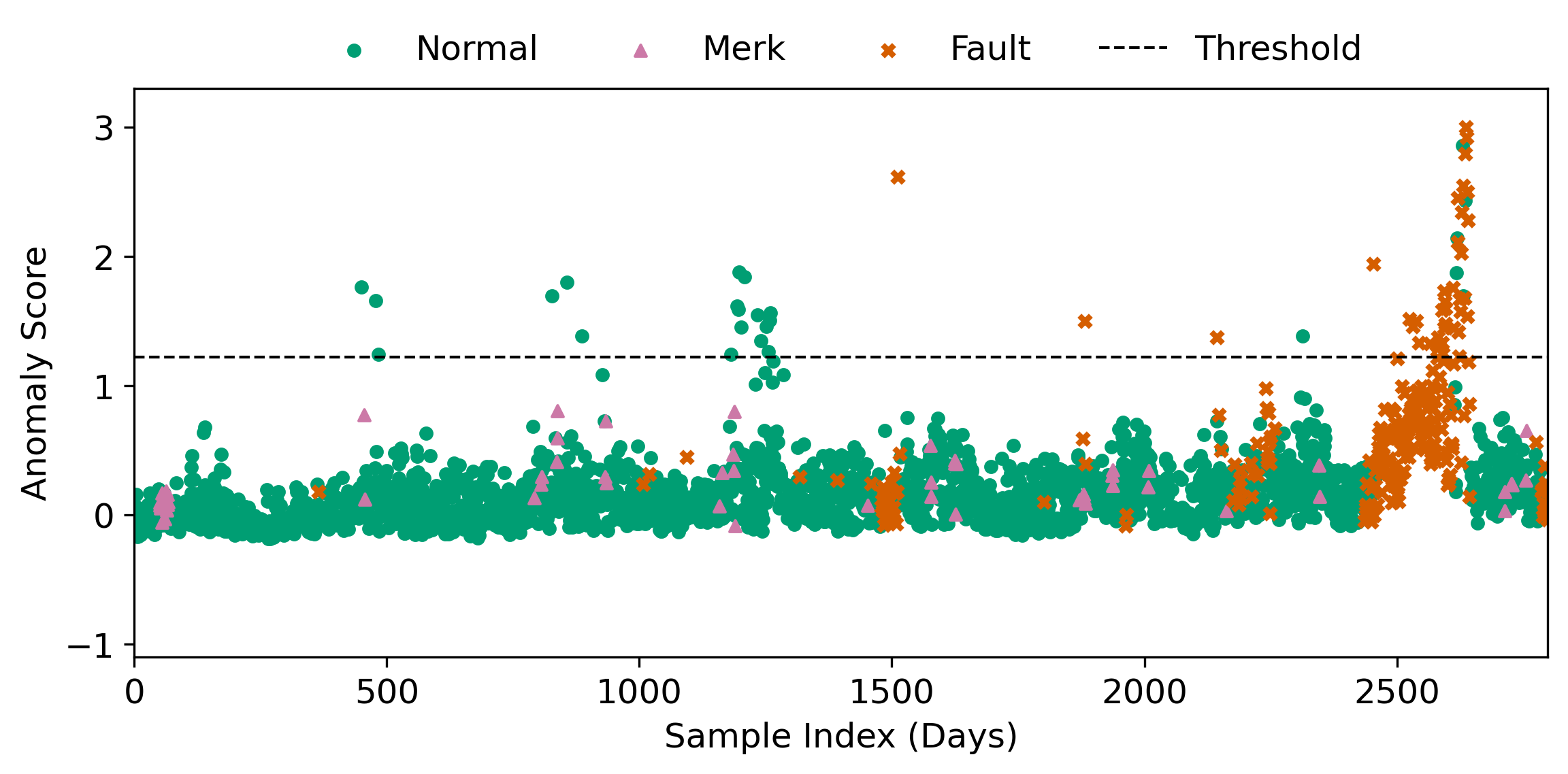}
    \caption{\texttt{Rescaled PCA-R}.}
  \end{subfigure}
  \caption{System~52 (F1 0.58 vs.\ 0.24). \texttt{MVE-$\beta$-VAE} (left) shows a trend in the anomaly score, indicating a consistent degradation of the system, leading to the complete failure of the system. The same trend is not observable in the PCA model (right).}
  \label{fig:comparison_52}
\end{figure}

\Cref{fig:comparison_52} shows the qualitative early detection capabilities of \texttt{MVE-$\beta$-VAE}. As discussed in the main text we can see a trend in the anomaly score time series, indicating a degradation of the system. This trend is already present before the critical issues arise later in the lifetime of the system, potentially allowing for early interventions. Compared to that, the PCA model has a less clear evolution of the anomaly score.

\begin{figure}[pos=htp!]
  \centering
  \begin{subfigure}[b]{0.49\linewidth}
    \centering
    \includegraphics[width=\linewidth]{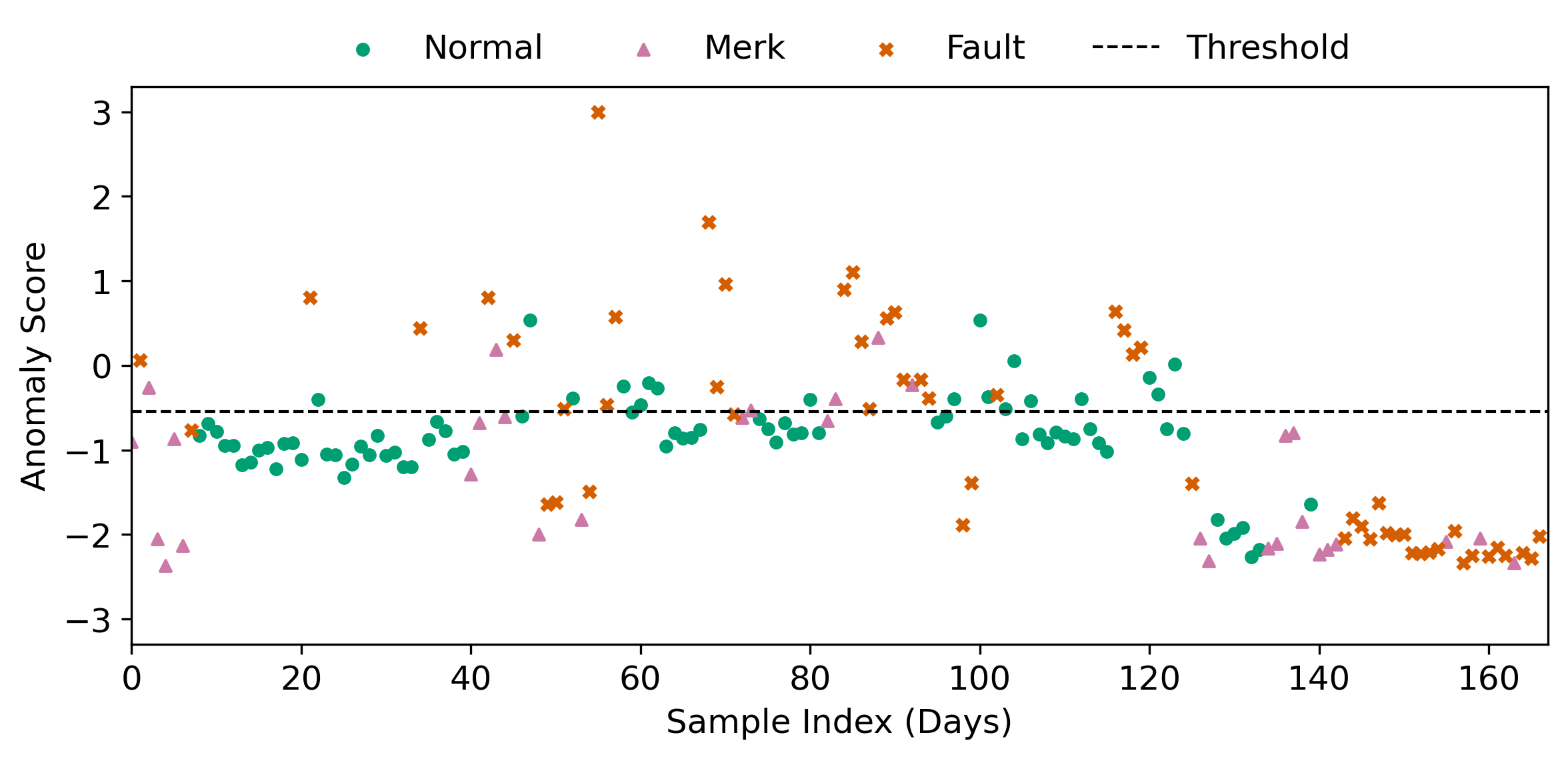}
    \caption{\texttt{MVE-$\beta$-VAE}.}
  \end{subfigure}
  \hfill
  \begin{subfigure}[b]{0.49\linewidth}
    \centering
    \includegraphics[width=\linewidth]{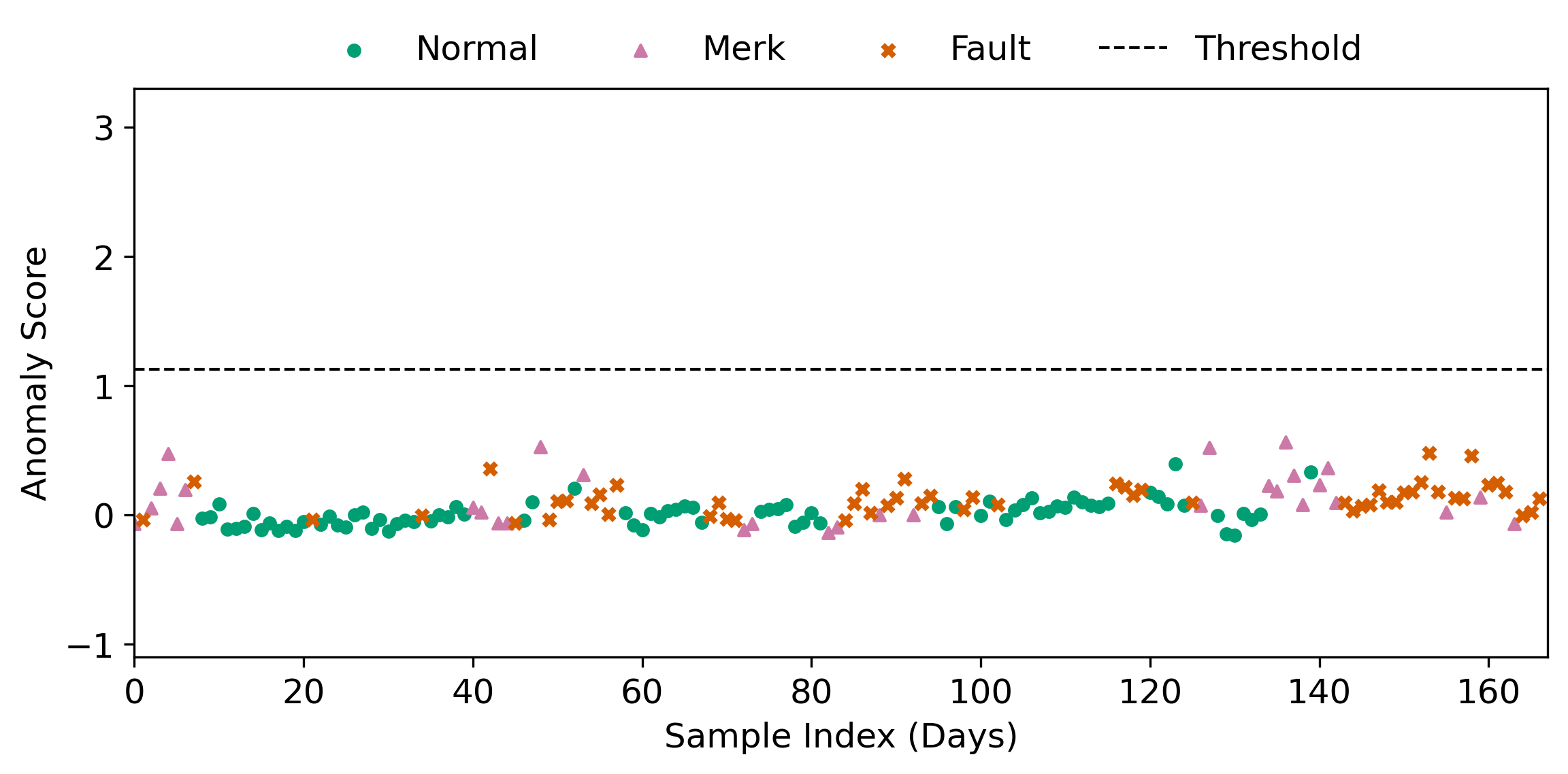}
    \caption{\texttt{Rescaled PCA-R}.}
  \end{subfigure}
  \caption{System~14 (F1 0.50 vs.\ 0.00). Qualitative analysis of the anomaly scores from \texttt{MVE-$\beta$-VAE} (left) correctly identifies the system as problematic. \texttt{Rescaled PCA-R} does not identify any faults here.}
  \label{fig:comparison_14}
\end{figure}
\begin{figure}[pos=htp!]
  \centering
  \begin{subfigure}[b]{0.49\linewidth}
    \centering
    \includegraphics[width=\linewidth]{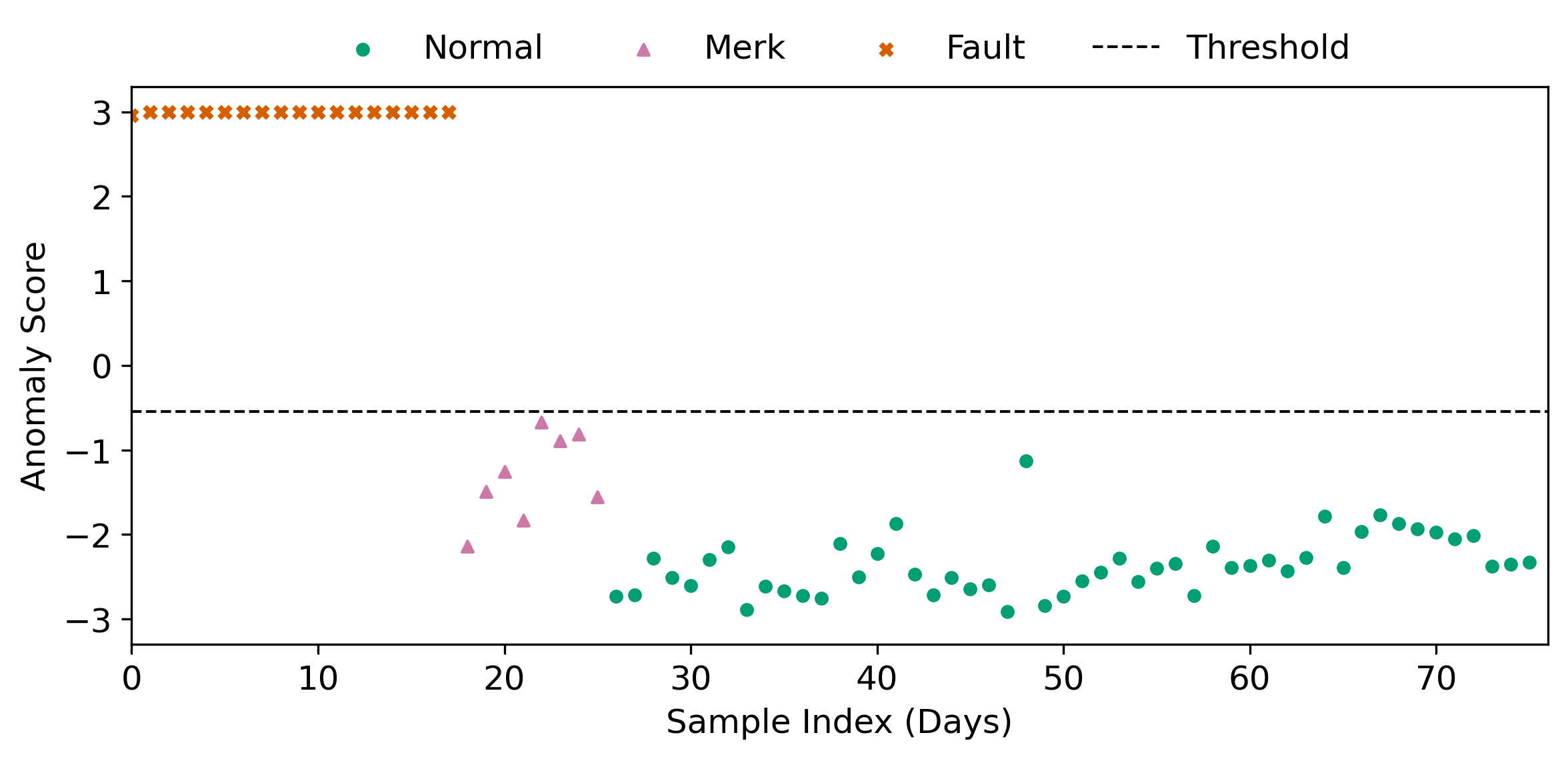}
    \caption{\texttt{MVE-$\beta$-VAE}.}
  \end{subfigure}
  \hfill
  \begin{subfigure}[b]{0.49\linewidth}
    \centering
    \includegraphics[width=\linewidth]{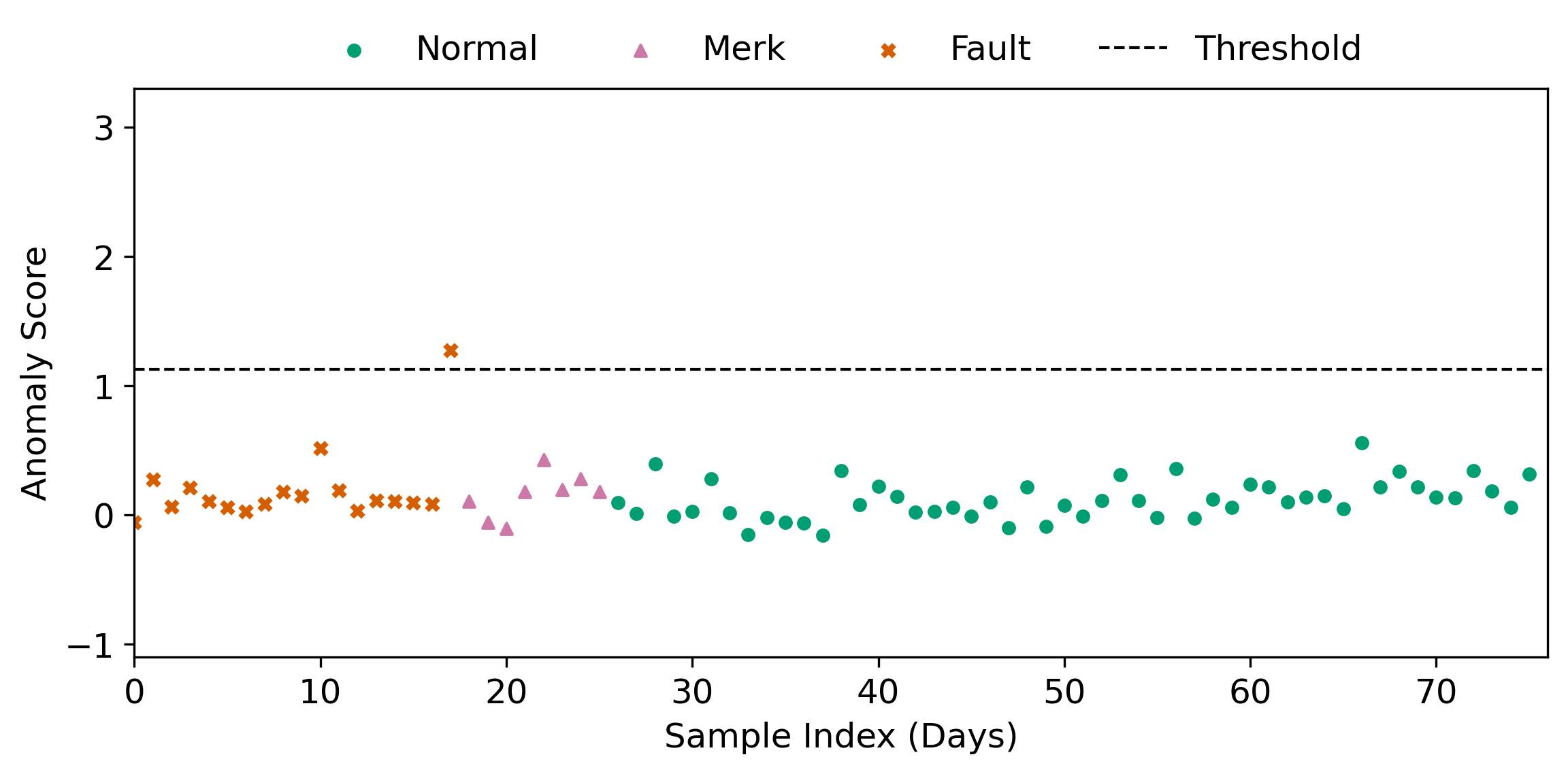}
    \caption{\texttt{Rescaled PCA-R}.}
  \end{subfigure}
  \caption{System~107 (F1 1.00 vs.\ 0.11). \texttt{MVE-$\beta$-VAE} (left) correctly identifies the system as faulty, while the PCA model (right) only detects the last day of the fault indication.}
  \label{fig:comparison_107}
\end{figure}
\begin{figure}[pos=htp!]
  \centering
  \begin{subfigure}[b]{0.49\linewidth}
    \centering
    \includegraphics[width=\linewidth]{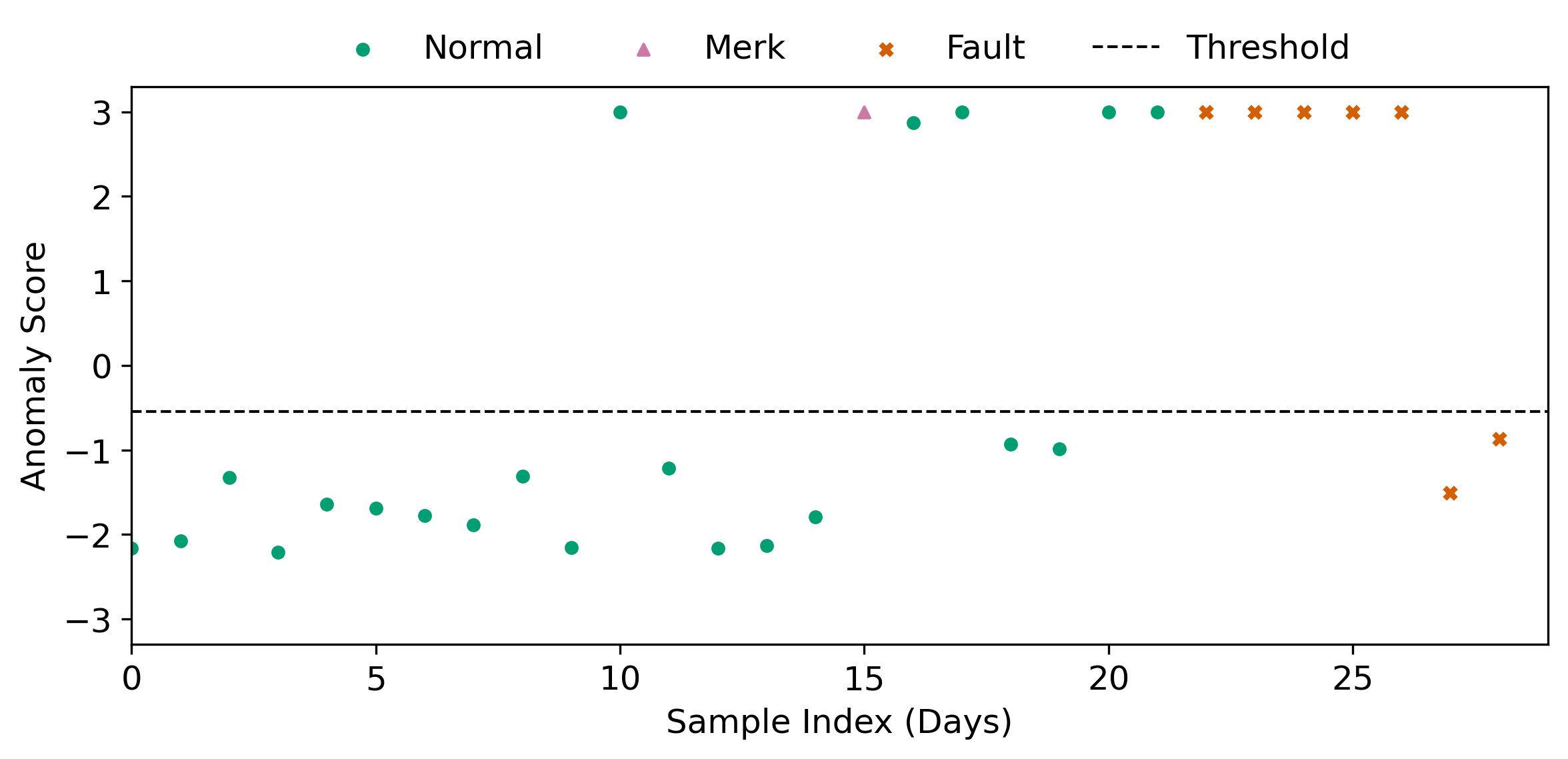}
    \caption{\texttt{MVE-$\beta$-VAE}.}
  \end{subfigure}
  \hfill
  \begin{subfigure}[b]{0.49\linewidth}
    \centering
    \includegraphics[width=\linewidth]{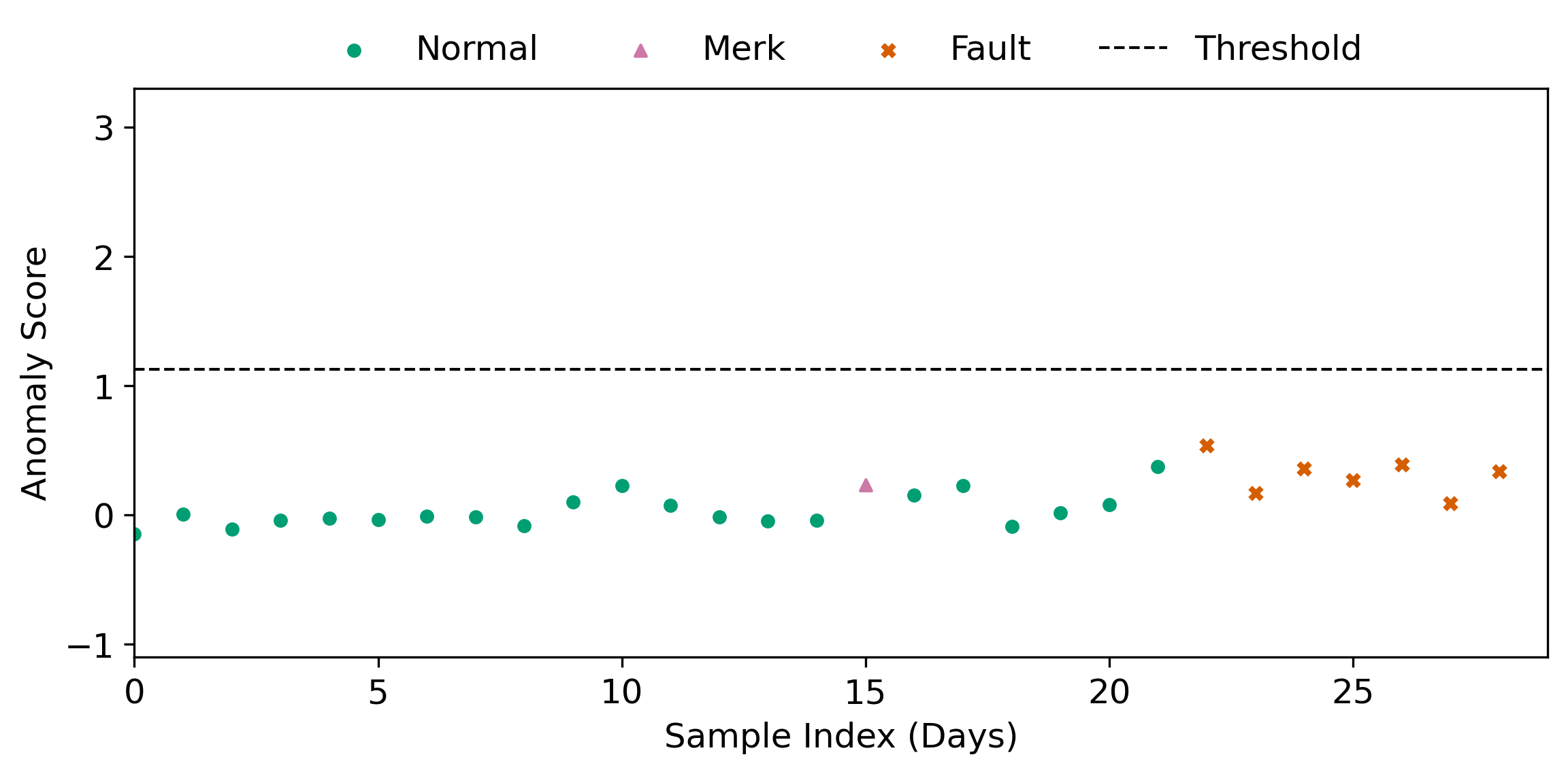}
    \caption{\texttt{Rescaled PCA-R}.}
  \end{subfigure}
  \caption{System~119 (F1 0.56 vs.\ 0.00). \texttt{MVE-$\beta$-VAE} (left) correctly identifies the system as faulty, with the first fault indications occurring before the system's internal fault indication flags the system as faulty. The PCA model (right) does not detect any system problem.}
  \label{fig:comparison_119}
\end{figure}

In \Cref{fig:comparison_107}, \Cref{fig:comparison_119}, and \Cref{fig:comparison_14} the PCA model fails to detect the faults present in the system, while \texttt{MVE-$\beta$-VAE} correctly identifies the systems as faulty.

\begin{figure}[pos=htp!]
  \centering
  \begin{subfigure}[b]{0.49\linewidth}
    \centering
    \includegraphics[width=\linewidth]{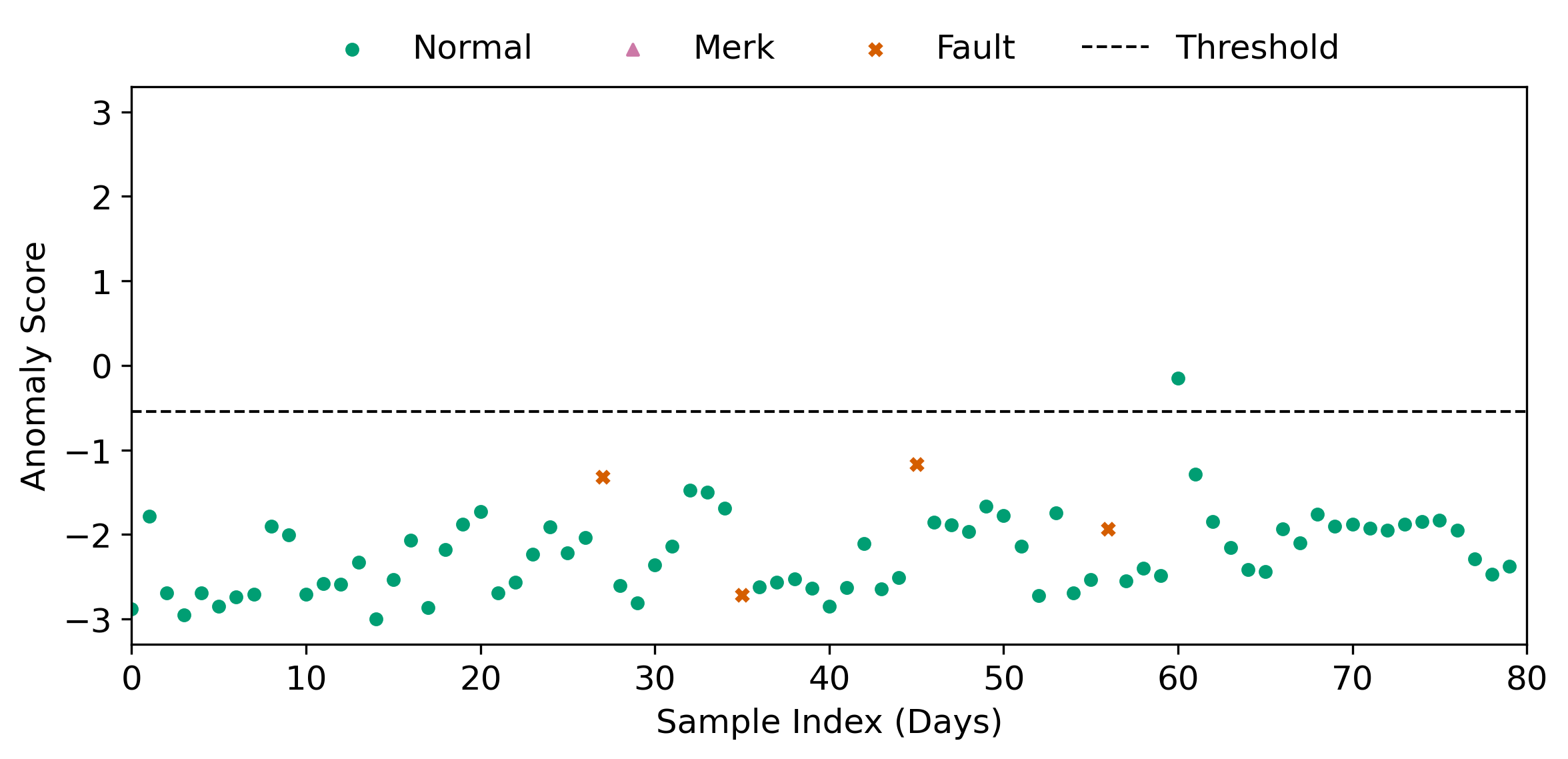}
    \caption{\texttt{MVE-$\beta$-VAE}.}
  \end{subfigure}
  \hfill
  \begin{subfigure}[b]{0.49\linewidth}
    \centering
    \includegraphics[width=\linewidth]{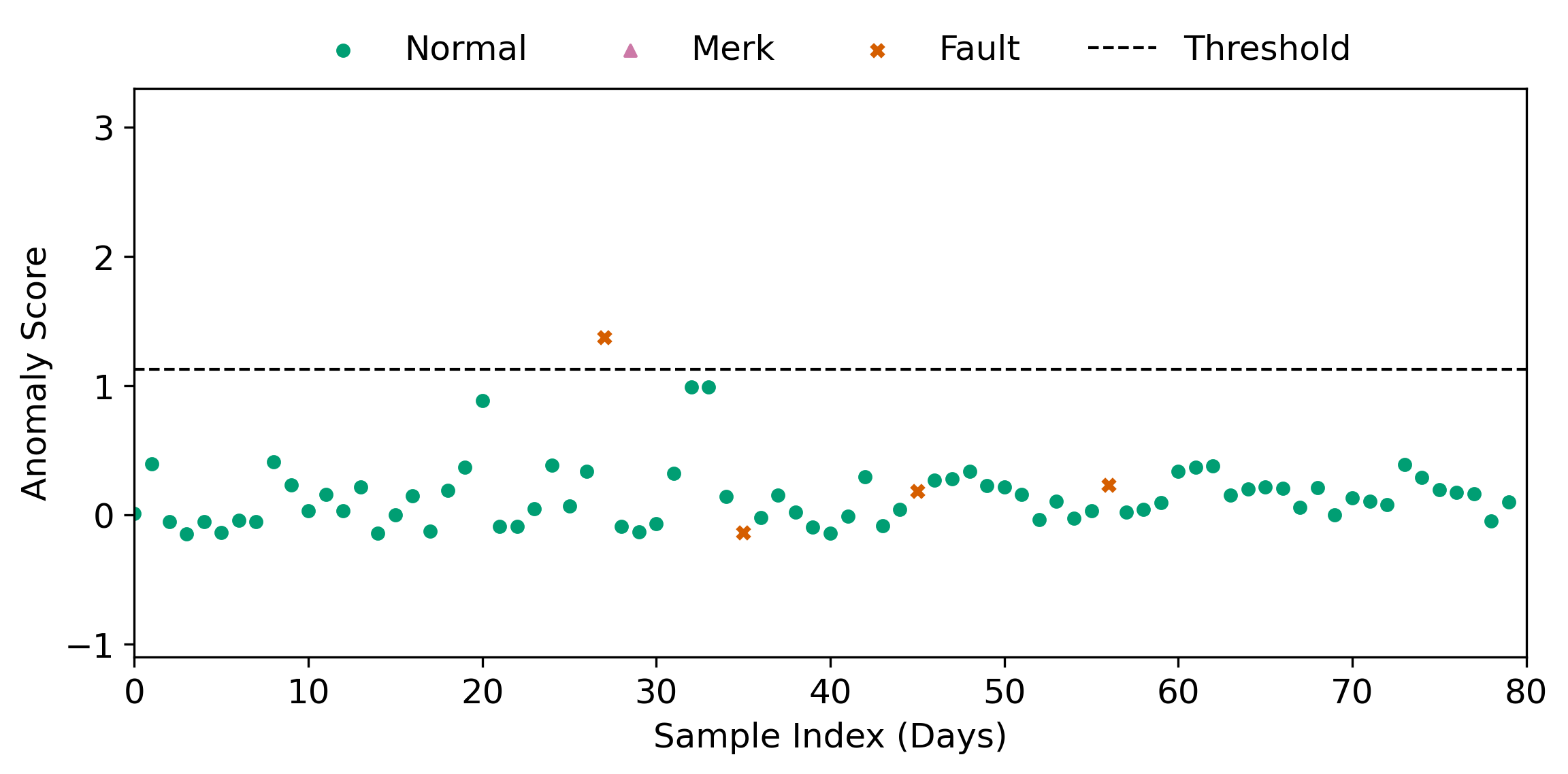}
    \caption{\texttt{Rescaled PCA-R}.}
  \end{subfigure}
  \caption{System~179 (F1 0.00 vs.\ 0.40). The record contains only a few isolated fault days. \texttt{Rescaled PCA-R} detects a single one of them, while \texttt{MVE-$\beta$-VAE} detects none; with so few fault days, this single detection produces a large F1 gap.}
  \label{fig:comparison_179}
\end{figure}
\begin{figure}[pos=htp!]
  \centering
  \begin{subfigure}[b]{0.49\linewidth}
    \centering
    \includegraphics[width=\linewidth]{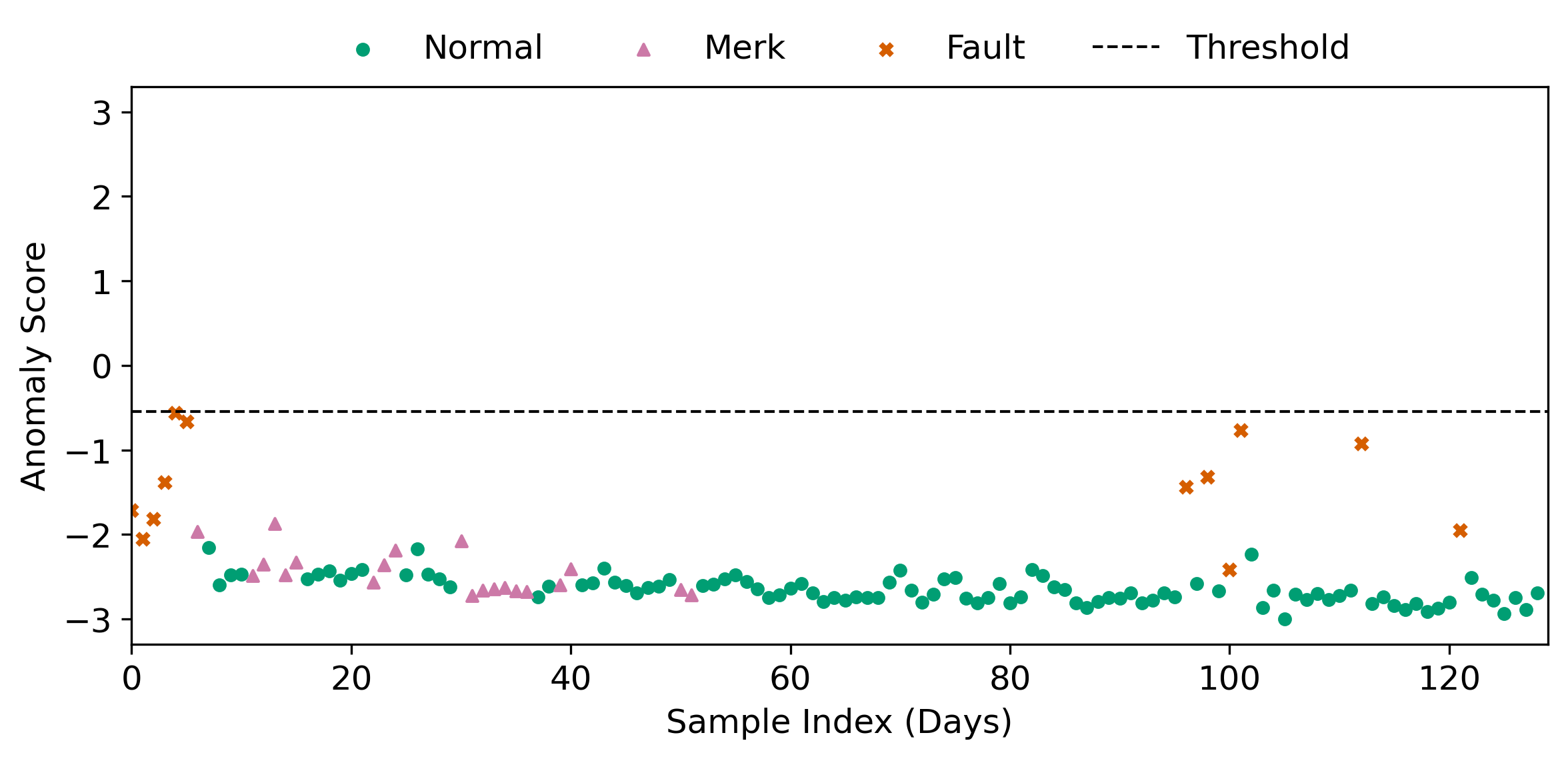}
    \caption{\texttt{MVE-$\beta$-VAE}.}
  \end{subfigure}
  \hfill
  \begin{subfigure}[b]{0.49\linewidth}
    \centering
    \includegraphics[width=\linewidth]{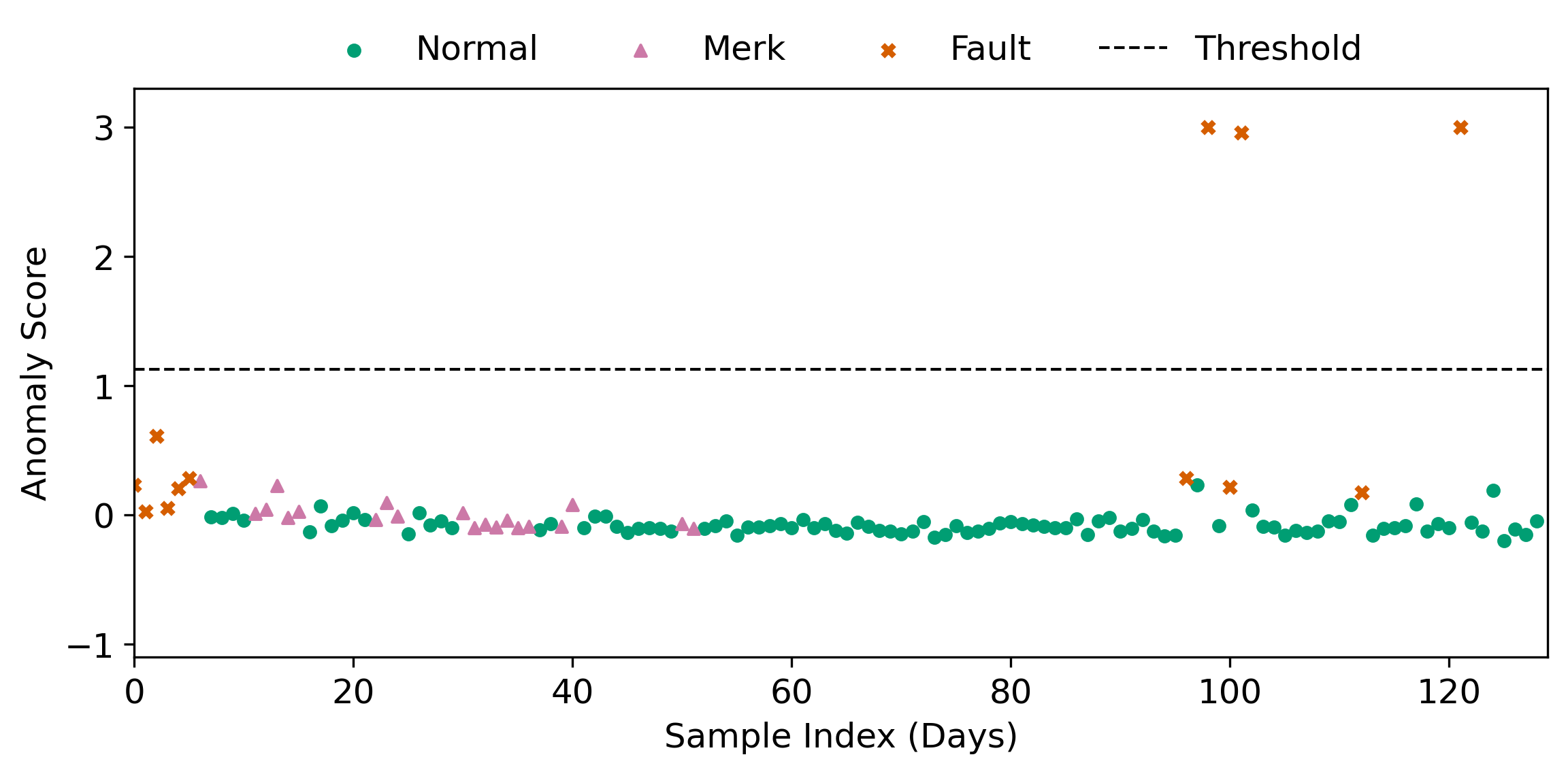}
    \caption{\texttt{Rescaled PCA-R}.}
  \end{subfigure}
  \caption{System~36 (F1 0.00 vs.\ 0.40). \texttt{Rescaled PCA-R} detects three days of the later fault period decisively, while \texttt{MVE-$\beta$-VAE} assigns anomaly scores slightly below the threshold in both fault periods, at the beginning and end of the record.}
  \label{fig:comparison_36}
\end{figure}
\begin{figure}[pos=htp!]
  \centering
  \begin{subfigure}[b]{0.49\linewidth}
    \centering
    \includegraphics[width=\linewidth]{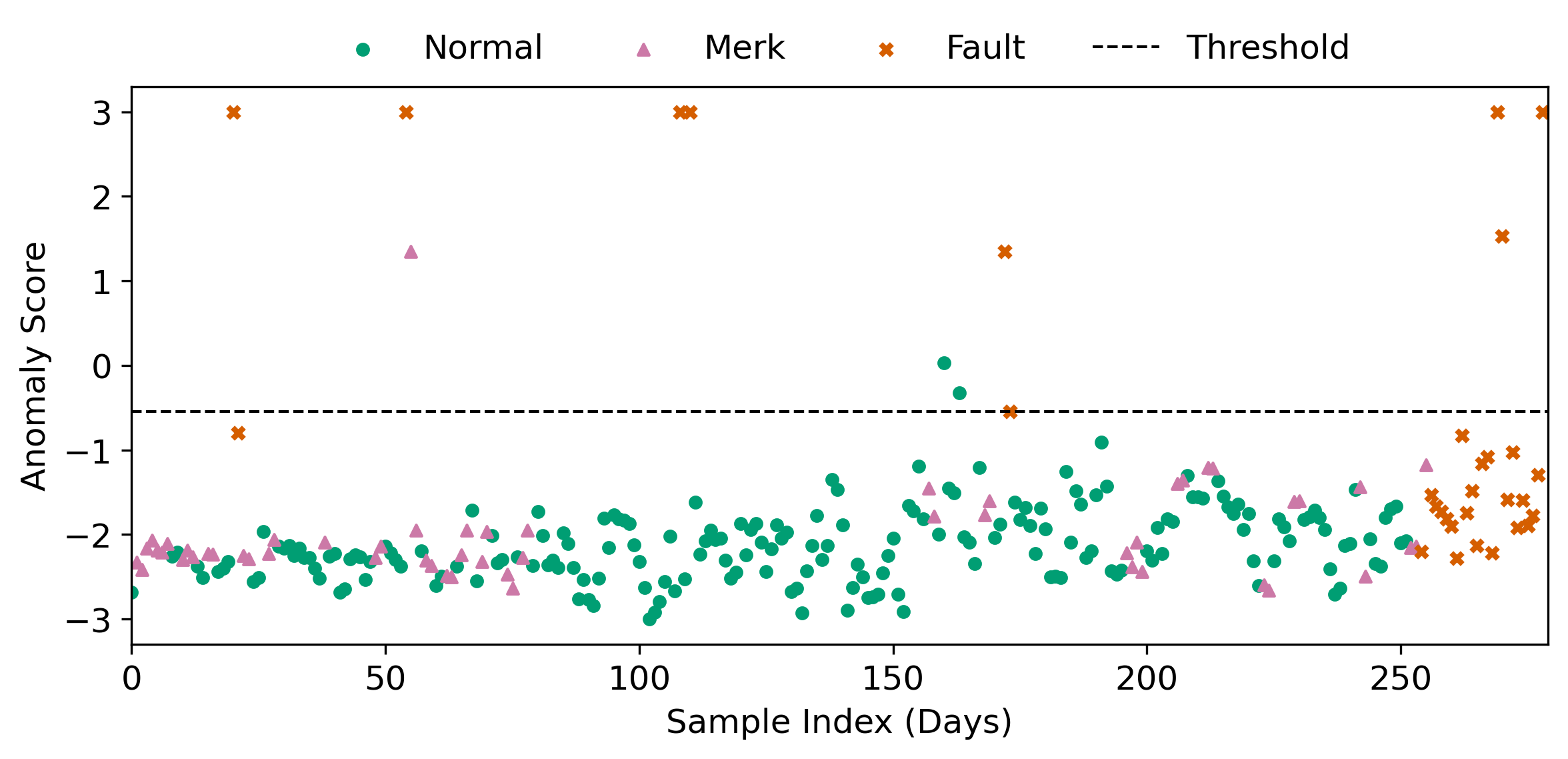}
    \caption{\texttt{MVE-$\beta$-VAE}.}
  \end{subfigure}
  \hfill
  \begin{subfigure}[b]{0.49\linewidth}
    \centering
    \includegraphics[width=\linewidth]{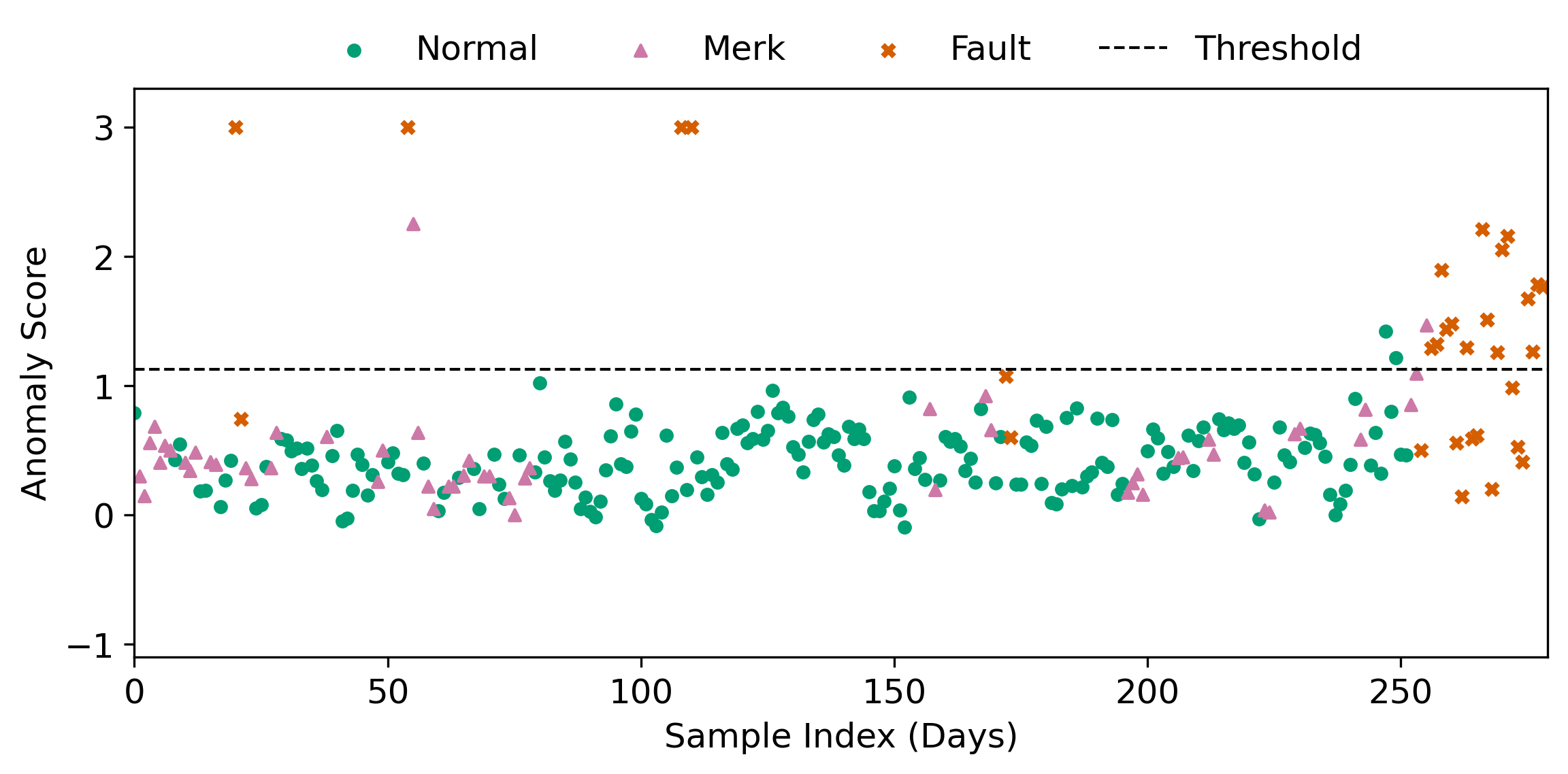}
    \caption{\texttt{Rescaled PCA-R}.}
  \end{subfigure}
  \caption{System~45 (F1 0.38 vs.\ 0.70). Both models detect the intermittent faults. \texttt{Rescaled PCA-R} additionally separates a large part of the fault period at the end of the record, which \texttt{MVE-$\beta$-VAE} misses at the operating threshold.}
  \label{fig:comparison_45}
\end{figure}
\begin{figure}[pos=htp!]
  \centering
  \begin{subfigure}[b]{0.49\linewidth}
    \centering
    \includegraphics[width=\linewidth]{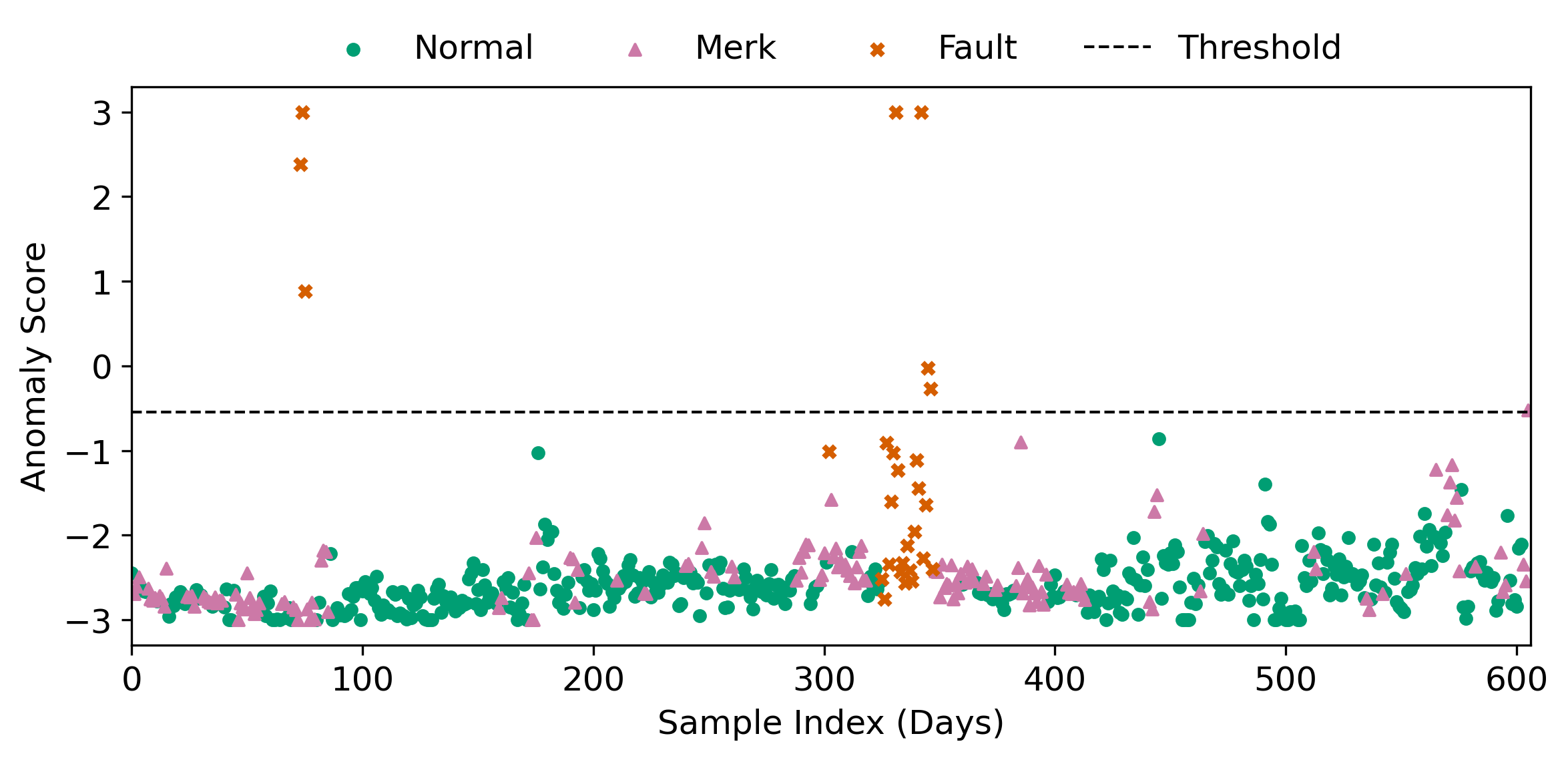}
    \caption{\texttt{MVE-$\beta$-VAE}.}
  \end{subfigure}
  \hfill
  \begin{subfigure}[b]{0.49\linewidth}
    \centering
    \includegraphics[width=\linewidth]{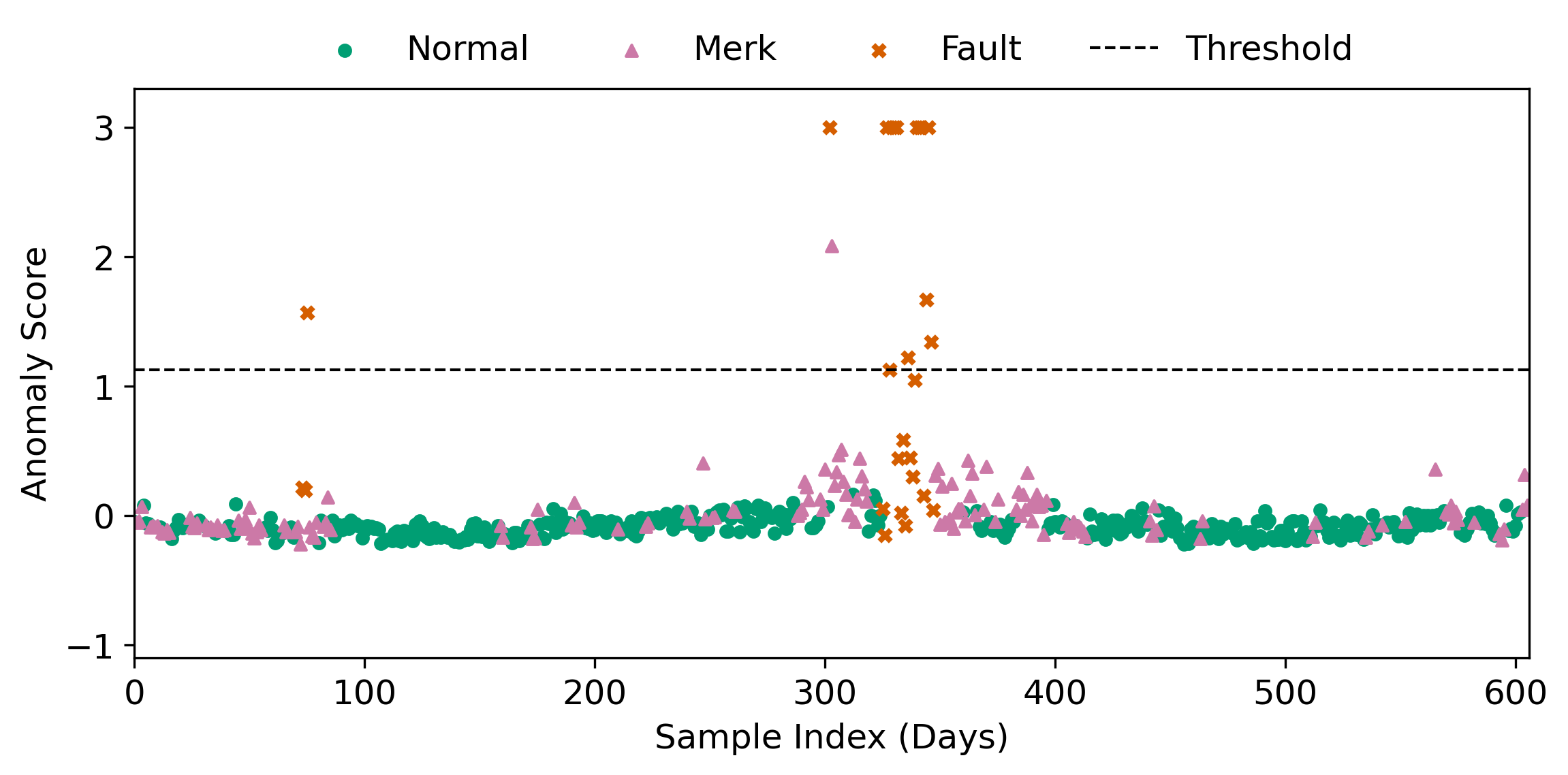}
    \caption{\texttt{Rescaled PCA-R}.}
  \end{subfigure}
  \caption{System~31 (F1 0.40 vs.\ 0.67). Both models detect the early fault cluster around day~80, but \texttt{Rescaled PCA-R} separates the main fault period around day~330 more strongly.}
  \label{fig:comparison_31}
\end{figure}

Conversely, \Cref{fig:comparison_179,fig:comparison_36,fig:comparison_45,fig:comparison_31} show the four systems with the largest F1 difference in favour of \texttt{Rescaled PCA-R} ($+0.40$, $+0.40$, $+0.32$, and $+0.27$). These are the systems furthest below the diagonal in \Cref{fig:system_wise_comparison}. In Systems~179 and~36, \texttt{MVE-$\beta$-VAE} detects no fault day at its operating threshold, mirroring the systems above the diagonal on which \texttt{Rescaled PCA-R} fails entirely. In System~179, the fault days are sparse and isolated within a short record, and \texttt{Rescaled PCA-R} detects a single one of them. In System~36, \texttt{Rescaled PCA-R} detects three days of the later fault period decisively. Here \texttt{MVE-$\beta$-VAE} barely misses the detection of both fault clusters as its operating threshold is slightly too high. In System~45, both models detect the intermittent faults, but \texttt{Rescaled PCA-R} additionally separates a large part of the fault period at the end of the record. In System~31, both models detect the early fault cluster around day~80, while \texttt{Rescaled PCA-R} separates the main fault period around day~330 more strongly.

\FloatBarrier

\section{Ablation of the BNLL Parameter $\beta_L$}
\label{sec:appendix_betal}
\begin{figure}[pos=htp!]
  \centering
  \begin{subfigure}[b]{0.49\linewidth}
    \centering
    \includegraphics[width=\linewidth]{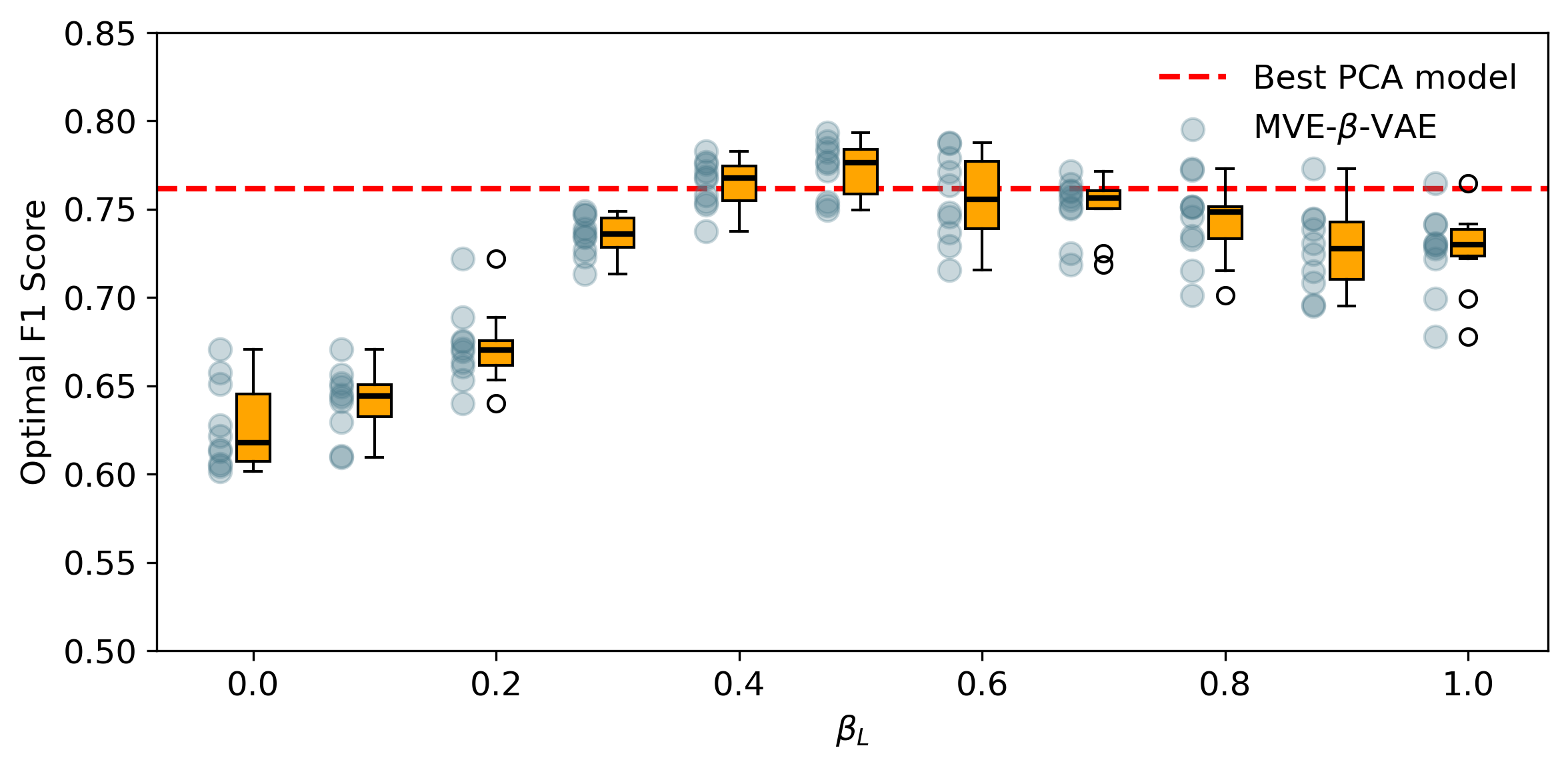}
    \caption{Optimal F1 score}
  \end{subfigure}
  \hfill
  \begin{subfigure}[b]{0.49\linewidth}
    \centering
    \includegraphics[width=\linewidth]{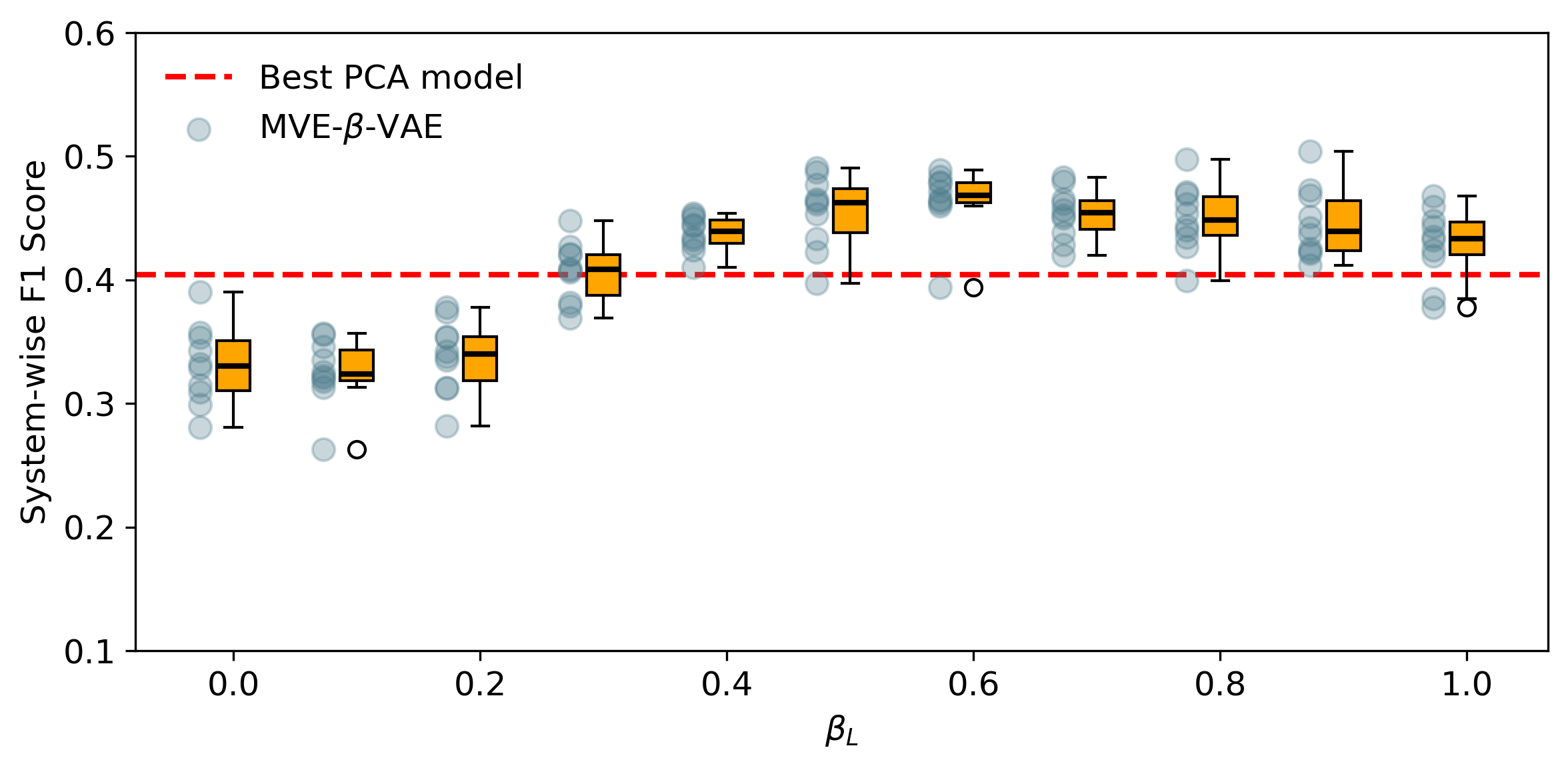}
    \caption{System-wise F1 score}
  \end{subfigure}
  \caption{Optimal F1 score (a) and system-wise F1 score (b) on the test set as a function of the BNLL parameter $\beta_L$, where $\beta_L = 0$ recovers the NLL and $\beta_L = 1$ the MSE. Each dot represents a single training run ($N = 10$ per setting). Boxes span the inter-quartile range with the median marked, whiskers extend to the furthest point within 1.5~inter-quartile ranges, and points beyond are shown individually. The dashed line marks the best PCA baseline.}
  \label{fig:betal_sweep}
\end{figure}
We swept the BNLL parameter $\beta_L$ from $0.0$ to $1.0$, where $\beta_L = 0$ recovers training with the standard NLL and $\beta_L = 1$ the MSE. All other hyperparameters were kept as in \Cref{tab:hyperparams}, and anomaly scores were computed with the standard NLL in all cases. Note that $\beta_L = 1$ recovers the MSE only for the mean prediction, while the model still produces heteroscedastic variance estimates. The NLL therefore remains a meaningful anomaly score in this case.

\Cref{fig:betal_sweep} reports the optimal F1 score and the system-wise F1 score on the test set. Detection performance peaks around $\beta_L = 0.4$--$0.6$ and degrades towards both extremes, with training on the pure NLL performing worst. This is consistent with the training pathologies of the NLL described by \citet{Seitzer2022Pitfalls} and supports their proposed default of $\beta_L = 0.5$, which we used in all other experiments. Only models trained with an intermediate $\beta_L$ match or exceed the best PCA baseline on the system-wise F1 score.

The sweep allows us to decouple the two roles of the heteroscedastic variance estimates: their influence on training, via the BNLL loss, and their influence on the anomaly score, via the NLL.
We evaluated the models trained with $\beta_L = 0.5$ (BNLL) and $\beta_L = 1$ (MSE) using both the NLL and the plain MSE as the anomaly score.
\Cref{tab:score_ablation} shows that both components contribute: NLL scoring yields the larger share of the improvement in system-wise F1 for both training variants, while BNLL training adds a further gain in both metrics. The combination used in the main paper performs best throughout.

Notably, the contribution of the variance estimates is concentrated in the system-wise F1 score. For $\beta_L = 1$, NLL scoring leaves the optimal F1 unchanged while improving the system-wise F1 by $0.09$. This reinforces our interpretation that heteroscedastic uncertainty estimation is central to generalisation across systems, as it renders the anomaly scores comparable between systems with different operating characteristics.

\begin{table}[htp]
  \centering
  \caption{Ablation decoupling the contribution of the heteroscedastic variance estimates at training time (BNLL, $\beta_L = 0.5$, vs.\ MSE-equivalent mean training, $\beta_L = 1$) and at scoring time (NLL vs.\ MSE anomaly score). We report mean $\pm$ standard deviation over $N = 10$ training runs. The combination used in the main paper is shown in bold. The models are taken from the $\beta_L$ sweep and were trained independently of the models in \Cref{tab:results1}.}
  \label{tab:score_ablation}
  \begin{tabular}{llcc}
    \toprule
    \textbf{Training}        & \textbf{Scoring} & \textbf{Optimal F1}          & \textbf{System-wise F1}      \\
    \midrule
    BNLL (\(\beta_L = 0.5\)) & \textbf{NLL}     & \(\mathbf{0.773 \pm 0.016}\) & \(\mathbf{0.455 \pm 0.030}\) \\
    BNLL (\(\beta_L = 0.5\)) & MSE              & \(0.741 \pm 0.011\)          & \(0.330 \pm 0.022\)          \\
    MSE (\(\beta_L = 1\))    & NLL              & \(0.726 \pm 0.024\)          & \(0.429 \pm 0.029\)          \\
    MSE (\(\beta_L = 1\))    & MSE              & \(0.726 \pm 0.010\)          & \(0.342 \pm 0.022\)          \\
    \bottomrule
  \end{tabular}
\end{table}

\section{Calibration of the Uncertainty Estimates}
\label{sec:appendix_calibration}

\begin{figure}[pos=htb]
  \centering
  \includegraphics[width=0.75\linewidth]{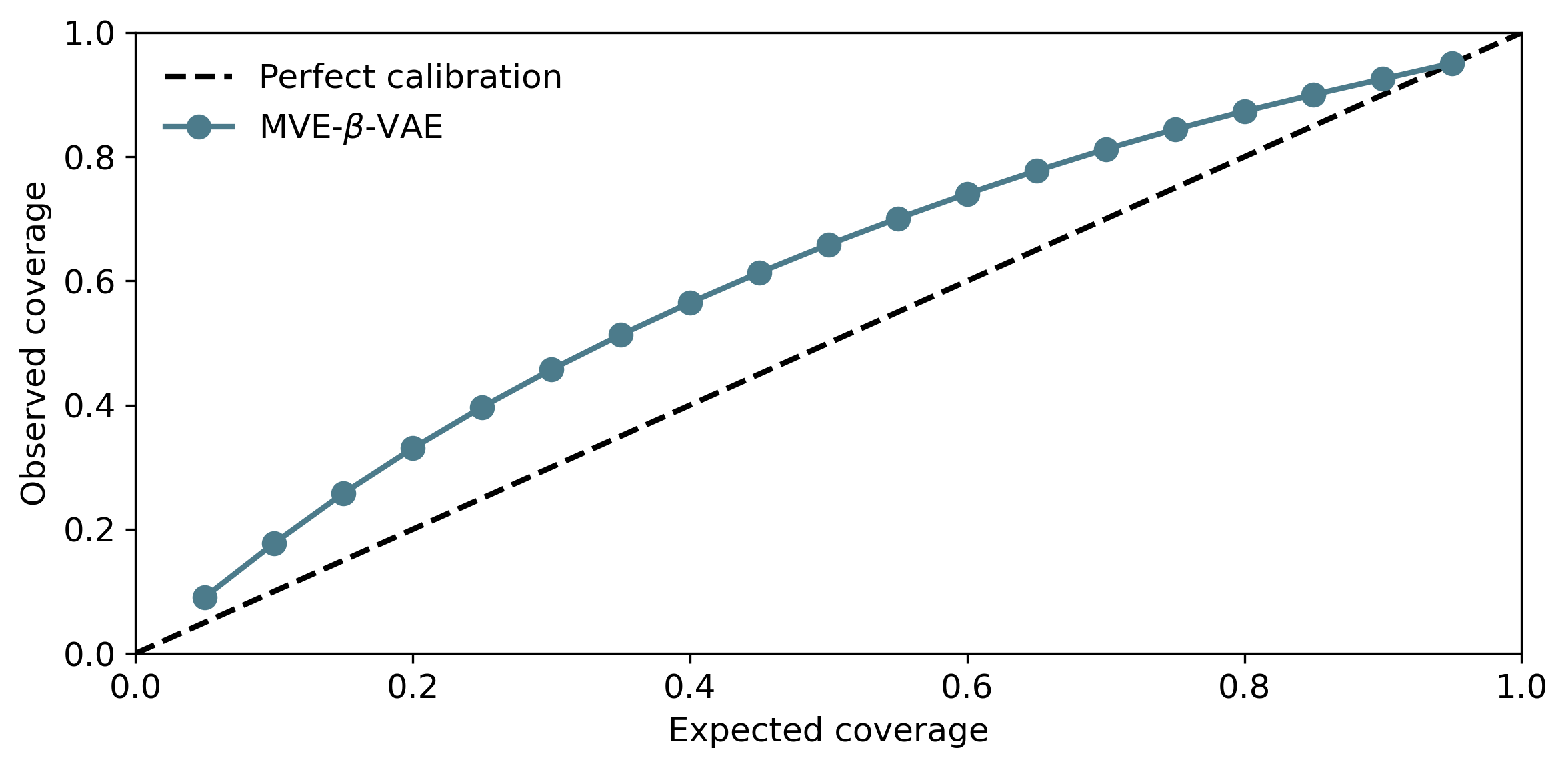}
  \caption{Calibration of the predicted variances on the held-out validation set. On the x-axis is the expected coverage under the Gaussian output distribution. The y-axis shows the observed coverage. The spread-skill ratio is 0.98.}
  \label{fig:calibration}
\end{figure}

We assessed the calibration of the predicted variances on the held-out nominal validation set. \Cref{fig:calibration} shows the observed coverage of the predicted central intervals against the expected coverage under the Gaussian output distribution.
The spread-skill ratio (SSR) on the validation set is $0.98$, indicating well-matched dispersion in aggregate. The coverage curve (\Cref{fig:calibration}) lies above the diagonal at intermediate levels: for the majority of datapoints, the predicted intervals are conservative, while a small fraction of large standardised residuals balances the aggregate ratio.
For anomaly detection, this mild conservatism is unproblematic, as the anomaly score depends on the relative consistency of the NLL between nominal and faulty days rather than on exact marginal coverage.

\section{Training Curves}
\label{sec:appendix_training}

\begin{figure}[pos=htb]
  \centering
  \includegraphics[width=\linewidth]{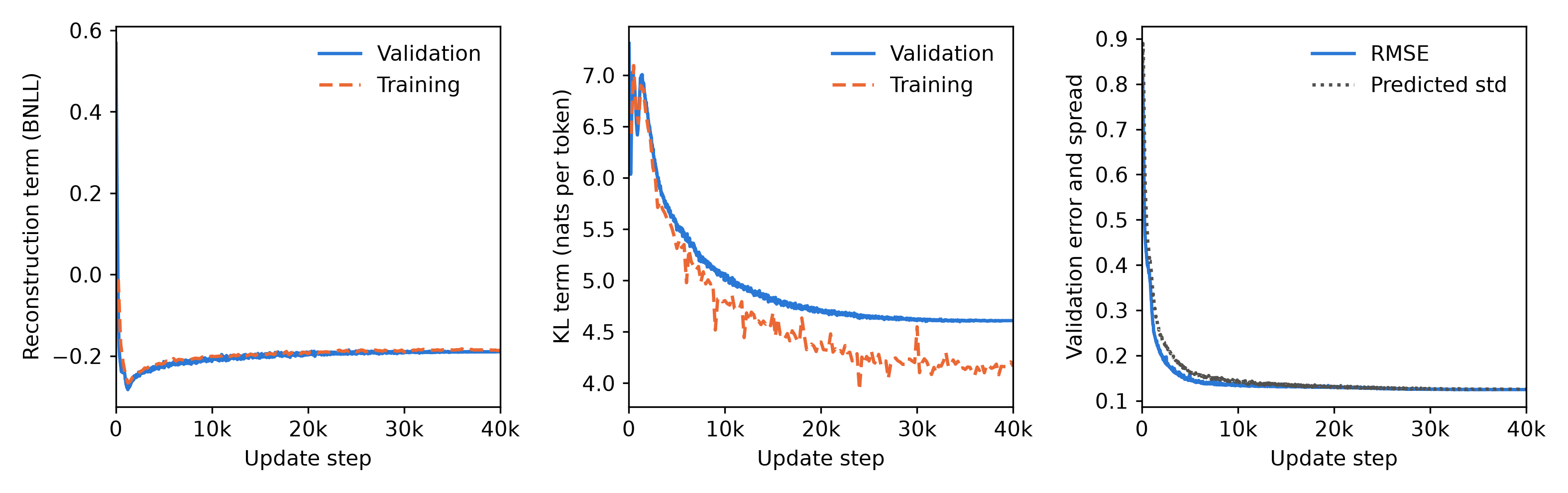}
  \caption{Training curves of a single \texttt{MVE-$\beta$-VAE} run with the hyperparameters of \Cref{tab:hyperparams}. Reconstruction term (left) and KL term (centre) of the loss on the training and validation set, and root mean squared error (RMSE) and root mean predicted variance on the validation set (right).}
  \label{fig:training_curves}
\end{figure}

\Cref{fig:training_curves} shows the training curves of \texttt{MVE-$\beta$-VAE}. The reconstruction term agrees closely between the training and validation set, with no signs of overfitting, and the KL term converges to $4.61$\,nats per token on the validation set (approximately $4.2$ on the training set).
As the BNLL is weighted by the predicted variance, its value is not directly comparable over the course of training: the reconstruction term increases after an early minimum, while the RMSE continues to decrease and the predicted variance follows it.

\section{Metrics Discussion}
\label{sec:appendix_metrics}
\begin{figure}[pos=htb]
  \centering
  \includegraphics[width=0.75\linewidth]{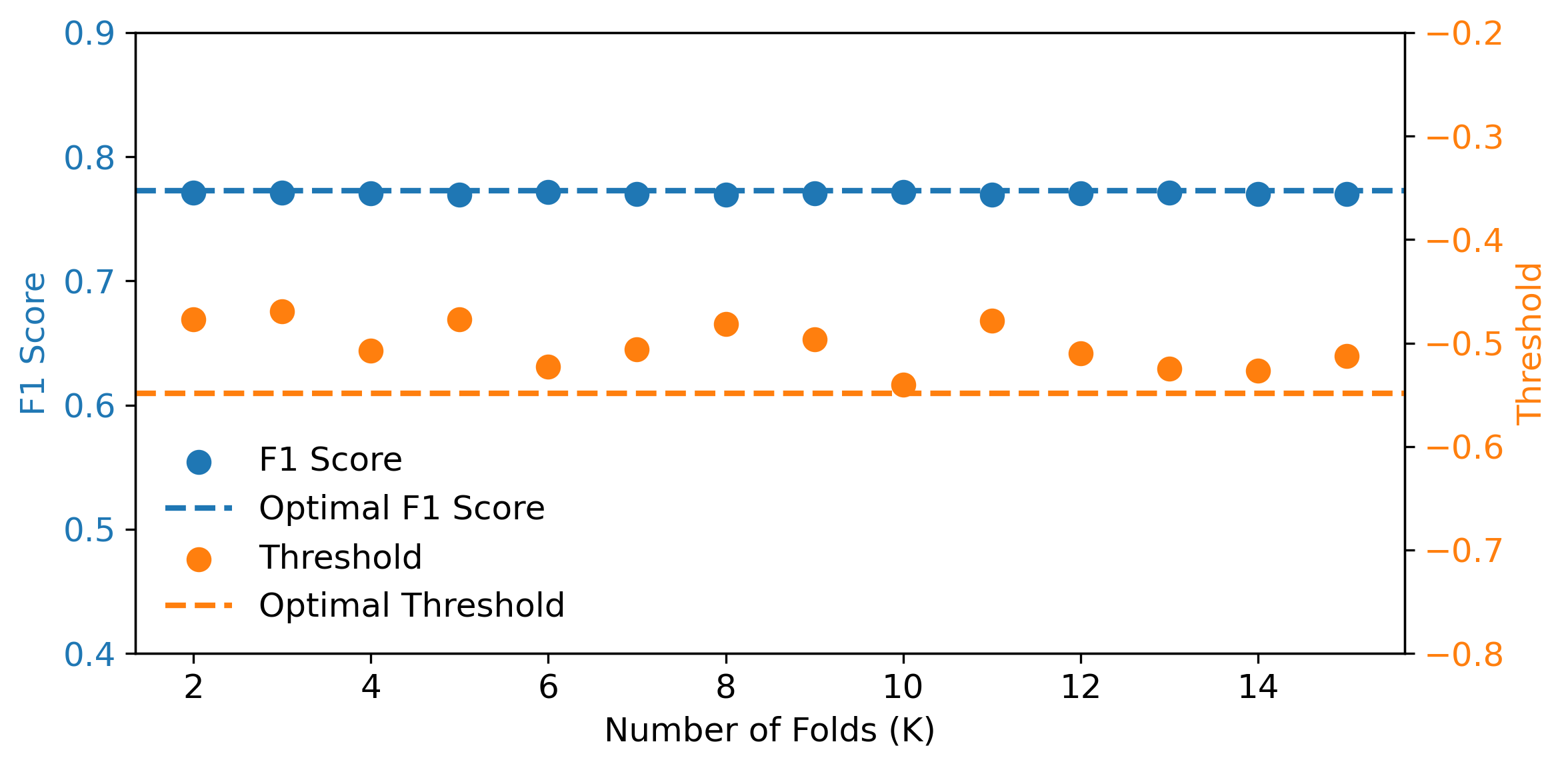}
  \caption{K-fold cross-validation for the F1 score. On the x-axis the number of folds is shown. The y-axis shows the average F1 score and the threshold. The dashed lines show the optimal threshold and optimal F1 score. The folds are partitioned at the day level.}
  \label{fig:kfold}
\end{figure}
As discussed in \Cref{sec:experimental_setup}, we evaluated our results using the optimal F1 score, the system-wise F1 score, the AUC-ROC and the AUC-PR. The optimal F1 score can be seen as an upper bound on the F1 score using a static threshold over the entire dataset. However, here the threshold is chosen as the optimum over the test set. To quantify how much this selection inflates the score, we estimated the threshold via K-fold cross-validation at the day level: the test days were randomly partitioned into $K$ folds, the threshold was estimated on $K-1$ folds and applied to the held-out fold, and the resulting F1 scores were averaged. \Cref{fig:kfold} shows the resulting F1 score for different K. As is clear in this plot, both the F1 score and the threshold do not substantially change using this method. This is why we opted to use the optimal F1 score metric, which is commonly used in time series anomaly detection.

Additionally, due to the strong imbalance in system length, the three standard metrics---optimal F1, AUC-ROC and AUC-PR---underestimate performance in small systems. Thus, they do not reflect generalisation capabilities of the models. However, the system-wise F1 score weights every system with the same weight, better representing generalisation capabilities.

\begin{table}[htp]
  \centering
  \caption{Confusion matrices of \texttt{MVE-$\beta$-VAE} (median seed) and \texttt{Rescaled PCA-R} at their respective optimal thresholds on the reannotated dataset.}
  \label{tab:confusion}
  \begin{tabular}{lrrrrrr}
    \toprule
    \textbf{Model}           & \textbf{TP} & \textbf{FP} & \textbf{FN} & \textbf{TN} & \textbf{Precision} & \textbf{Recall} \\
    \midrule
    \texttt{MVE-$\beta$-VAE} & 3060        & 597         & 1205        & 20634       & 0.84               & 0.72            \\
    \texttt{Rescaled PCA-R}  & 2914        & 473         & 1351        & 20758       & 0.86               & 0.68            \\
    \bottomrule
  \end{tabular}
\end{table}

\Cref{tab:confusion} shows the confusion matrices of \texttt{MVE-$\beta$-VAE} and \texttt{Rescaled PCA-R} at their respective optimal thresholds on the reannotated dataset. At this operating point, the two models trade precision and recall differently: \texttt{MVE-$\beta$-VAE} detects more faulty days (recall $0.72$ vs.\ $0.68$) at a slightly lower precision ($0.84$ vs.\ $0.86$).

\section{Percentile-based Thresholds}
\label{sec:appendix_percentile}

\begin{table*}[ht]
  \centering
  \caption{Anomaly detection performance under fixed thresholds calibrated only on nominal validation data (95th and 99th percentile of the validation anomaly scores of the respective model), evaluated on the reannotated dataset. We report the F1 score over all days and the system-wise F1 score at these thresholds. Values are the mean over all seeds ± the standard deviation; for non-stochastic methods no standard deviation is given. Baselines marked with $\dagger$ are forecasting-based deep learning models and forecast a single step. The best-performing model per metric and all models without a statistically significant difference to it are marked in bold (see \Cref{sec:experimental_setup}). Pairs of deterministic methods cannot be tested under this procedure.}
  \label{tab:percentile_results}
  \begin{tabular}{l l@{\,}r|l@{\,}r|l@{\,}r|l@{\,}r}
    \toprule
    \textbf{Method}                                   & \multicolumn{2}{@{}c|}{System-wise F1 (p95)} & \multicolumn{2}{c|}{F1 (p95)} & \multicolumn{2}{c|}{System-wise F1 (p99)} & \multicolumn{2}{c}{F1 (p99)}                                                                     \\
    \midrule
    \texttt{Unscaled PCA-R}                           & 0.36                                         &                               & 0.67                                      &                              & 0.27            &               & 0.63            &               \\
    \texttt{Rescaled PCA-R}                           & 0.42                                         &                               & \textbf{0.75}                             &                              & 0.33            &               & \textbf{0.75}   &               \\
    \texttt{MSE-$\beta$-VAE}                          & 0.44 ±                                       & 0.01                          & 0.62 ±                                    & 0.02                         & 0.33 ±          & 0.03          & 0.70 ±          & 0.03          \\
    \texttt{MVE-$\beta$-VAE}                          & \textbf{0.50} ±                              & \textbf{0.01}                 & 0.66 ±                                    & 0.03                         & \textbf{0.46} ± & \textbf{0.05} & \textbf{0.76} ± & \textbf{0.02} \\
    \midrule
    \texttt{HBOS\cite{goldstein2012histogram}}        & 0.18                                         &                               & 0.48                                      &                              & 0.09            &               & 0.35            &               \\
    \texttt{LOF\cite{breunig2000lof}}                 & 0.37                                         &                               & 0.44                                      &                              & 0.26            &               & 0.44            &               \\
    \texttt{KMeansAD\cite{yairi2001fault}}            & 0.32 ±                                       & 0.01                          & 0.50 ±                                    & 0.01                         & 0.23 ±          & 0.01          & 0.57 ±          & 0.01          \\
    \texttt{PCA\cite{paffenroth2018robust}}           & 0.30                                         &                               & 0.71                                      &                              & 0.19            &               & 0.65            &               \\
    \texttt{IForest\cite{liu2008isolation}}           & 0.21                                         &                               & 0.45                                      &                              & 0.20            &               & 0.50            &               \\
    \midrule
    \texttt{AnomalyTransformer\cite{xu_anomaly_2022}} & 0.22 ±                                       & 0.05                          & 0.30 ±                                    & 0.05                         & 0.13 ±          & 0.05          & 0.21 ±          & 0.07          \\
    \texttt{AutoEncoder\cite{sakurada2014anomaly}}    & 0.22 ±                                       & $<$ 0.01                      & 0.29 ±                                    & $<$ 0.01                     & 0.21 ±          & $<$ 0.01      & 0.29 ±          & $<$ 0.01      \\
    \texttt{CNN\cite{munir2018deepant}}$\dagger$      & 0.44 ±                                       & 0.01                          & 0.59 ±                                    & 0.02                         & 0.37 ±          & 0.02          & \textbf{0.75} ± & \textbf{0.01} \\
    \texttt{Donut\cite{tran2016distance}}             & 0.36 ±                                       & 0.02                          & 0.60 ±                                    & 0.01                         & 0.28 ±          & 0.02          & 0.67 ±          & 0.02          \\
    \texttt{LSTMAD\cite{malhotra2015long}}$\dagger$   & 0.37 ±                                       & 0.01                          & 0.58 ±                                    & 0.02                         & 0.34 ±          & 0.02          & 0.64 ±          & 0.02          \\
    \texttt{TimesNet\cite{wu2022timesnet}}            & 0.23 ±                                       & 0.01                          & 0.34 ±                                    & 0.01                         & 0.17 ±          & 0.01          & 0.29 ±          & 0.02          \\
    \texttt{TranAD\cite{tuli2022tranad}}              & 0.35 ±                                       & 0.02                          & 0.65 ±                                    & 0.04                         & 0.32 ±          & 0.02          & 0.67 ±          & 0.03          \\
    \texttt{USAD\cite{audibert_usad_2020}}            & 0.26 ±                                       & 0.01                          & 0.36 ±                                    & 0.02                         & 0.24 ±          & $<$ 0.01      & 0.36 ±          & 0.01          \\
    \bottomrule
  \end{tabular}
\end{table*}

The optimal and system-wise F1 scores reported in \Cref{tab:results1,tab:results2} rely on thresholds informed by labelled evaluation data. As an alternative to this, we evaluated the model on fixed thresholds calibrated on nominal data. For each model we took the 95th and 99th percentile of its anomaly scores on the nominal validation set and applied those as a threshold to the test data. \Cref{tab:percentile_results} reports the resulting F1 and system-wise F1 scores on the reannotated dataset.

The ranking is consistent with the main results, with \texttt{MVE-$\beta$-VAE} achieving the best system-wise F1 at both percentiles. For some models, the scores at the 95th percentile threshold even exceed those in \Cref{tab:results1}. This indicates that a similar performance to that in \Cref{tab:results1} is attainable even without any labelled data, as long as clean, nominal training data exists. A proper choice of percentile is a deployment parameter which would need to be tuned on operational feedback and essentially trades detection rate against false alarms.
\end{document}